\documentclass[pre,twocolumn,showpacs,superscriptaddress]{revtex4}
\usepackage{graphicx,amsfonts}
\usepackage{amsmath}

\begin{document}
\newcommand{\be}{\begin{equation}}
\newcommand{\ee}{\end{equation}}
\newcommand{\ba}{\begin{aligned}}
\newcommand{\ea}{\end{aligned}}
\newcommand{\bi}{\begin{itemize}}
\newcommand{\ei}{\end{itemize}}
\newcommand{\bn}{\begin{enumerate}}
\newcommand{\en}{\end{enumerate}}
\newcommand{\rv}{\vec{r}}
\newcommand{\vv}{\vec{v}}
\newcommand{\Fv}{\vec{F}}
\newcommand{\im}{\item}
\newcommand{\bc}{\begin{center}}
\newcommand{\ec}{\end{center}}
\newcommand{\ta}{\hat {t}}
\newcommand{\nn}{\hat{n}}
\newcommand{\nij}{\nn _{ij}}
\newcommand{\tij}{\ta _{ij}}
\newcommand {\defeq}{\stackrel{\mbox{\tiny def}}{=}}
\newcommand {\qq}   {\,\qquad}
\newcommand {\qqp}  {\,;\qquad}
\newcommand {\tab} [1] {table\,\ref{tab-#1}}
\newcommand{\ave}[1]{\langle #1 \rangle}
\newcommand{\norm}[1]{\left|\left| #1 \right|\right|}
\newcommand{\abs}[1]{\left| #1 \right|}
\newcommand {\scbox} [1] {\mbox{\scriptsize #1}}
\newcommand{\disty}{\displaystyle}
\newcommand {\as}{\text{I}}
\newcommand {\RR}  { {\mbox{\tiny RR}  } }
\newcommand {\nRR} { {\mbox{\tiny no RR} } }
\newcommand {\fmin}  { F_{\mbox{\tiny min} } }
\newcommand {\fmax}  { F_{\mbox{\tiny max} } }
\newcommand {\fave}  { F_{\mbox{\tiny ave} } }
\newcommand {\phim}  { \Phi_{\mbox{\tiny max} } }
\newcommand{\degr}{^\circ}
\let\eps = \varepsilon
\newcommand {\Hyp} { {\mbox{$_2$F$_1$}} }
\newcommand{\ww}[1]{\underline{\underline{{\bf #1}}}}
\newcommand{\sti }{\ww{K}}
\newcommand{\stia }{\sti^{(1)}}
\newcommand{\stib }{\sti^{(2)}}
\newcommand{\trs}[1]{\hspace{.05em}\,^{\bf T}\hspace{-.1em}\ww{#1}}
\newcommand{\rig }{\ww{G}}
\newcommand{\rigt }{\trs{G}}
\newcommand{\fext}{{\bf F}^{\text{ext}}}
\newcommand{\abso}[1]{\vert #1 \vert}
\newcommand{\hhh}{\textsc{h}}
\newcommand{\kkk}{\textsc{k}}
\newcommand{\kk}{{\bf k}}
\title{Computer simulation of model cohesive powders:\\
Plastic consolidation, structural changes and elasticity under isotropic loads.}

\author{F.~A.~Gilabert}
\email{gilav@us.es} \affiliation{Faculty of Physics, University of
Seville, Avda. Reina Mercedes s/n, 41012 Seville, Spain.}

\author{J.-N.~Roux}
\affiliation{ Universit\'e Paris-Est, Institut Navier,\\
Laboratoire des Mat\'eriaux et des Structures du G\'enie Civil\footnote{LMSGC
is a joint laboratory depending on Laboratoire Central des Ponts et Chauss\'ees, \'Ecole Nationale des
Ponts et Chauss\'ees and Centre National de la Recherche
Scientifique.}, 2 All\'ee Kepler, Cit\'e Descartes,
77420 Champs-sur-Marne, France.}

\author{A.~Castellanos}
\affiliation{Faculty of Physics, University of Seville, Avda. Reina
Mercedes s/n, 41012 Seville, Spain.}

\date{\today}

\begin{abstract}
The quasistatic behavior of a simple 2D model of a cohesive powder under isotropic
loads is investigated by Discrete Element simulations.
We ignore contact plasticity and focus on the effect of geometry and collective rearrangements on the material behavior.
The loose packing states, as assembled and characterized in a previous
numerical study [Gilabert, Roux and Castellanos, Phys. Rev. E {\bf 75}, 011303 (2007)],
are observed, under growing confining pressure $P$, to undergo important
structural changes, while solid fraction $\Phi$ irreversibly increases (typically, from 0.4--0.5 to 0.75--0.8).
The system state goes through three stages, with different forms of the \emph{plastic consolidation curve}, \emph{i.e.},
$\Phi$ as a function of the growing reduced pressure $P^*=Pa/F_0$, defined with adhesion force $F_0$ and
grain diameter $a$.
In the low-confinement regime (I), the system undergoes negligible plastic compaction, and its structure
is influenced by the assembling process.
In regime II the material state is
independent of initial conditions, and the
void ratio varies linearly with $\log P$ [i. e. $\Delta(1/\Phi)=\lambda \Delta (\log P^*)$],
as described in the engineering literature. Plasticity index $\lambda$ is reduced in the presence of
a small rolling resistance (RR).
In the last stage of compaction (III),
$\Phi$ approaches an asymptotic, maximum solid fraction $\phim$,
as a power law, $\phim -\Phi \propto (P^*)^{-\alpha}$, with $\alpha\simeq 1$, and
properties of cohesionless granular packs are gradually retrieved.
Under consolidation, while the range $\xi$ of fractal density correlations decreases, force patterns
reorganize from self-balanced clusters to force chains, with correlative evolutions of force distributions, and elastic moduli increase by a large amount.
Plastic deformation events correspond to very small changes in the network topology,
while the denser regions tend to move like rigid bodies. Elastic properties are dominated by the bending of thin junctions in loose systems.
For growing RR those tend to reduce to particle chains, the folding of which, rather than tensile ruptures, controls plastic compaction.
\end{abstract}

\maketitle

\section{Introduction \label{sec:intro}}
Cohesive granular materials are present in many natural or industrial processes, the understanding of which requires studies of their rheology under
small confining pressures, when tensile intergranular forces play a major role.  In such cases cohesive materials
exhibit specific features that do not exist in cohesionless grain assemblies, such as the ability to form stable structures at low density and
the sensitivity to stress \emph{intensity}, as opposed to stress direction.
Macroscopic constitutive laws and phenomenological tools have been developed and used in several engineering fields:
mechanics of cohesive soils (clays and silts)~\cite{Wood90,MIT93,BiHi93,Atkinson93}, 
metallic powder processing~\cite{Poquillon02}, 
modeling and treatment of ceramic powders~\cite{Cooper1962,Reed1995,Falgon2005,PhDGustavo},
handling of xerographic toners~\cite{C05}.
One simple material is the assembly of wet beads~\cite{PiCa97,GTH03,Fournier05}, in which some microscopic observations
are possible~\cite{GTH03,Fournier05}. However, wet grain packs are only slightly less dense than dry ones, and do not enable the study of loose
structures obtained with powders.
In general, the behavior of materials under proportional load (oedometric or isotropic compression) is characterized by the
consolidation curve, which describes the irreversible compaction under growing stress~\cite{Wood90}. Density can  increase
by factors of 3 or 4 under growing load.

Although numerical simulations have been widely used for several decades~\cite{CS79} to investigate microscopic mechanisms and classify mechanical properties of granular systems,
studies of cohesive materials are still far less common, and almost exclusively limited to dense materials. Thus, the effects of
capillary cohesion in wet sand or bead packs have been simulated~\cite{RiEYRa06,RiRaEY06}, as well as the compaction of
ceramic and metallic powders~\cite{Kong00,MaBo03,Martin03,MBS03,Martin04,Nam04,Martin06} to states of very high density,
or the behavior in shear tests of 2D dense cohesive packs with plastic deformation of
contacts~\cite{LU05,TTLKHB07}. Loose structures formed by particles packed under gravity and stabilized thanks to adhesion have been simulated~\cite{DYZY06}.
Of particular relevance to the present study, among the very scarce numerical studies of \emph{loose} packings~\cite{Misra96} stabilized by cohesion and of their collapsing under growing loads, are the works by Bartels, Kadau, Wolf~\emph{et al.}~\cite{KBBW02,KBBW03,GUKWK05,WoUnKaBr05}
on the oedometric compression of granular assemblies with initial low densities. This research group studied
a dynamical compression regime, and observed a shock wave propagating through the sample. Shear flows of cohesive granular materials have also been
simulated~\cite{BrGrLaLe05,AaSu06,RoRoWoNaCh06,Rognon08}.

In a previous article~\cite{GRC07}, hereafter referred to as paper I, we studied by numerical simulation the assembling process, the structure
and the force patterns of a model, two-dimensional (2D) cohesive granular material in loose equilibrium configurations.
We now investigate the mechanical behavior of the same model granular
material in isotropic compression and pressure cycles, as well as the evolution of various characteristics of intermediate equilibrium states as plastic
compaction proceeds.

As in paper I, we keep the external pressure as the main control parameter. The adhesive strength $F_0$ in contacts sets a force scale in the material behavior,
and hence (in 2D) the reduced pressure, defined as
\be
P^* = \frac{aP}{F_0},
\label{eq:defPstar}
\ee
in which $a$ is a typical grain diameter, is a crucial dimensionless state parameter. The main objective
of the present paper is the study of the process by which, as pressure is increased, cohesion-dominated loose structures, for which $P^*\ll 1$,
get irreversibly compacted as $P^*$ increases until pressure dominates ($P^*\gg 1$). Such a compaction was numerically observed \emph{e.g.,} in Ref.~\cite{WoUnKaBr05}.
However, our approach produces homogeneous, isotropic, equilibrium configurations under varying load
and is therefore apt to provide more detailed information about the connections between macroscopic constitutive laws and microstructural or micromechanical features.


The present paper is self-contained and can be understood without reading paper I.
%
%
%
%
A summarized description of the material properties and of the initial configurations (studied
in paper I) is provided in Section~\ref{sec:model}. The macroscopic material response in isotropic compression,
with the possible influence of the initial state properties, is studied in Section~\ref{sec:macro}. Then, various microscopic aspects of the consolidation process are
investigated in the sequel: density correlations (with their fractal behavior over some length scale~\cite{GRC07})
are investigated in Section~\ref{sec:geom}, force networks and force distributions are dealt with in Section~\ref{sec:forces}, while Section~\ref{sec:elasmod} focuses on elastic
moduli. Section~\ref{sec:compmicro} discusses qualitatively some microscopic aspects of the consolidation behavior.
The final section, part~\ref{sec:conc}, summarizes the results and suggests directions for future work. Sections~\ref{sec:geom} and~\ref{sec:forces} can, at first, be read
independently from each other. The same remark applies to Sections~\ref{sec:elasmod} and~\ref{sec:compmicro}.
\section{Model material and simulation procedures \label{sec:model}}
\subsection{Definitions and basic equations\label{sec:param}}
The material and the simulation method are identical to those of paper I~\cite{GRC07},
which the reader might refer to for additional technical details, and for a physical discussion of some of the model ingredients.
For the sake of completeness,
we however provide a summarized description below. The contact law is an elaboration of the often employed spring-dashpot model with Coulomb friction,
in which two additional ingredients are introduced: an attractive force and, possibly, some resistance to rolling at contacts.
The model material is a 2D assembly of disks with diameters uniformly
distributed between $a/2$ and $a$, enclosed in a rectangular cell with periodic boundary conditions
in both directions. Both lengths $L_1$, $L_2$ defining the cell size and shape are variable, and
satisfy equations of motion designed to impose given values of diagonal stress components $\sigma_1 = \sigma_2 = P$.
Stresses are controlled by a variant of the Parrinello-Rahman method~\cite{PARA81}. In equilibrium, both diagonal stress
components $\sigma_{\alpha}$, ($\alpha=1,2$), with the convention that tensile stresses are negative,
are given by the standard formula ($A$ is the sample surface area):
\be
\sigma _{\alpha}= \frac{1}{A}\sum_{1\le i<j\le N} F_{ij}^{(\alpha)} r_{ij}^{(\alpha)}.
\label{eq:stress}
\ee
In~\eqref{eq:stress}, the r.h.s. sum runs over all interacting pairs $i,j$ among the $N$ disks of the system, ${\bf F}_{ij}$ is the force transmitted from
grain $i$ to its neighbor $j$ and vector ${\bf r}_{ij}$ points from the center of $i$ to the center of $j$ (with the suitable nearest image convention to account
for periodicity). The implementation of stress-controlled simulations is such that the cell length $L_\alpha$ along direction $\alpha$ increases or decreases
if $\sigma _{\alpha}$ is larger (respectively: smaller) than its prescribed value.

As usual in molecular dynamics applied to granular materials (also known as the ``discrete element method'')
particles have rigid body kinematics and their motion is governed by Newton's equations.
\subsection{Interaction law}
Grains interact with forces of elastic, adhesive, frictional and viscous origins.
The static part of the normal component $F_N^{ij}$ of the force transmitted by grain $i$ to its neighbor $j$ is a function of $h_{ij}$, the distance separating
disk perimeters.
A negative $h_{ij}$ means that the grains overlap, in which case they repel each other with a normal elastic force $F_N^{e,ij}=-K_N h_{ij}$. This force
vanishes whenever $h_{ij}>0$. (Overlap $h_{ij}<0$ is, of course, a numerical representation of the physical contact deflection).
The repulsive elastic force is supplemented with an attractive term $F_N^{a,ij}$, equal to $-F_0$ for contacting disks ($h_{ij}<0$). $F_N^{a,ij}$ has a finite
range $D_0$, fixed to $10^{-3}a$, and varies linearly between $-F_0$ and zero as $ h_{ij}$ grows from $0$ to $D_0$.
$F_0$ is the maximum tensile force a contact might support without breaking off. The normal contact law thus introduces a force scale,
and a dimensionless parameter, the \emph{stiffness parameter}, $\kappa \equiv aK_N / F_0$.  $\kappa$ characterizes the
amount of elastic deflection $h_0$ under contact force $F_0$,
relative to grain size $a$ ($h_0/a = \kappa^{-1}$).
$\kappa$ is set to a large value, $\kappa=10^5$, so that the elastic deflections in contacts remain so small that they can be neglected in comparison
to all other length scales in the problem (including interstices between neighbors~\cite{iviso1}). The packing geometry can be regarded as that of an assembly
of rigid grains (as formally dealt with in the ``contact dynamics'' simulation method used in~\cite{WoUnKaBr05}).

To the static contributions $F_N^e$ and $F_N^a$ to the normal force we add a viscous damping term
opposing the relative normal velocity of $i$ and $j$ when the disks touch ($h_{ij}<0$), corresponding to a constant,
positive normal coefficient of restitution $e_N$ in binary collisions if $F_0$ is set to zero. $e_N$ is set to a low value, $e_N=0.015$ in our simulations.
In the presence of attractive forces the apparent
restitution coefficient in a collision will depend on the initial
relative velocity. For small kinetic energies the particles will eventually stick to each other.
The minimum receding velocity for two particles of unit mass (the unit mass is chosen equal to the mass of a disk of
diameter $a$) to separate is $V^*\sqrt{2}$, with
\be
V^*=\sqrt{F_0D_0}. \label{eq:vsep}
\ee

The elastic tangential force in contact $i,j$, $F_T^{ij}$, is to be evaluated incrementally.
In case of no tangential sliding, it varies linearly with the relative
tangential displacement at the contact point, involving a tangential stiffness constant, $K_T$.
In the case of sliding, which occurs when the elastic law would cause $F_T^{ij}$ to pass one of the Coulomb bounds
$\pm \mu F_N^{e,ij}$, then $F_T^{ij}$ stays equal to $\pm F_N^{e,ij}$. The relative
tangential displacement at the contact point involves displacements of disk centers and rotations. The Coulomb condition introduces
the \emph{friction coefficient}, $\mu$.
It should be pointed out that it applies to the elastic repulsive part of the normal force only.
Thus, a pair of contacting grains with $h_{ij}$ equal to $F_0/K_N = h_0$, the equilibrium distance, such that the sum of elastic and adhesive
terms vanishes, can transmit a tangential force $F_T$ such that $\abso{F_T}\le \mu F_0$.
(The importance of this feature of the contact law for collective properties macroscopic behavior of
particle assemblies was stressed in paper I for isotropic, static states, and in Ref.~\cite{RoRoWoNaCh06}
in steady-state shear flows). All simulations reported here were carried out with $\mu=0.5$.

We studied the influence of \emph{rolling resistance} (RR) at contacts, which is modeled as in~\cite{TOST02}. Two additional parameters are necessary:
a \emph{rolling spring constant}, $K_R$, with dimension of a moment, expressing proportionality between relative rotation and rolling moment (i. e., a torque
concentrated at the contact point), as long as the rolling friction threshold is not reached; and a \emph{rolling friction coefficient}, $\mu_R$
with the dimension of a length, setting the maximum absolute value of the rolling moment $\Gamma_R$ to $\mu_RF_N^e$, proportional to the elastic
part of the normal force. The implementation of this rolling law is analogous to that of the tangential one, with the rolling moment
and the relative rotation respectively replacing the tangential force and the relative tangential displacement. A contact for which
the total normal force is equal to zero in equilibrium, with $F_N^e= K_N h_0= F_0$, may transmit a rolling moment $\Gamma_R$ with $\abso{\Gamma_R}\le \mu_R F_N^e$.
Since point contacts do not transmit torques, the rolling resistance stems from the irregularity of grain surface. Two contacting grains
touch each other, in general, by two points (in 2D), which are separated by some microscopic distance $l$ that is characteristic of the particle shape. $\mu_R$ should be
proportional to $l$, and $K_R$ proportional to $l^2$. We set $\mu_R = \mu l$ and $K_R = K_N l^2$, with, in most calculations with RR, $l = a/100$.

Table~\ref{tab:param} summarizes the values of parameters used in most simulations, in dimensionless form.
\begin{table}[!htb]
\centering
\begin{tabular}{ccccccc}\cline{1-7}
$\mu$  & $e_N$ &  {\Huge \strut} $\kappa$& ${\disty \frac{K_T}{K_N}\strut }$ & ${\disty \frac{D_0}{a}}$ &
${\disty \frac{K_R}{K_Na^2}}$& ${\disty \frac{\mu_R}{a}}$  \\ \hline\hline
$\ 0.5\ $&$\ 0.015\ $&$10^5$&$1$&$\ 10^{-3}\ $&$10^{-4}$&$\ \ 0$ or $0.005$\\ \hline
\end{tabular}
\caption{\label{tab:param} Values of dimensionless model
parameters used in most simulations.}
\end{table}
Some calculations were also performed with larger RR (up to $l=a$, $\mu_R=0.5a$).
%
\subsection{Initial states\label{sec:init}}
\begin{table*}[!htb]
\centering
\begin{tabular}{cccccc}
\hline
Sample type& No cohesion &Type 1& \multicolumn{3}{c}{Type 2} \\
\hline
$N$ & 1400   & 1400 & 1400 & 5600 & 10976 \\
Number of samples  &4& 4& 5& 3& 1   \\
Lowest pressure & $P/K_N=10^{-5}$ & $P^*=0.01$&\multicolumn{3}{c}{ $P^*=0.01$}\\
$\Phi$ (no RR)           &$0.811\pm 0.001$ & $0.723\pm 0.001$  &\multicolumn{3}{c}{ $0.472\pm 0.008$}\\
$\Phi$ (RR)           & $0.805\pm 0.002$ & $0.688\pm 0.001$  & \multicolumn{3}{c}{$0.524\pm 0.008$ }   \\
\hline
\end{tabular}
\caption{Set of granular samples used as initial equilibrated configurations in simulations of isotropic compression (with material parameters of Table~\ref{tab:param}).
\label{tab:list}}
\end{table*}
In paper I, two extreme cases were studied in the assembling stage of cohesive packings under low $P^*$. First, an $N$-particle sample of hard-disk fluid is prepared at
solid fraction $\Phi_I$ in a fixed cell. Then, in type 1 systems, velocities are set to zero
and the external pressure control is started, until an equilibrium is reached under $P^*=0.01$. The other procedure, by which type 2 samples are prepared, is meant to
represent the opposite situation, in which aggregation is much faster than compression. Thus, while the cell size is fixed and the solid fraction stays equal to $\Phi_I$,
grains are attributed random (Maxwell-distributed) velocities and left to interact and aggregate until all $N$ of them join to form one unique cluster. The system
is then equilibrated at $P^*=0$, and compressed to $P^*=0.01$. To limit the influence of dynamical effects, the strain rate is requested not to exceed a maximum
value $\dot\epsilon_{\text{max}}$ during compression. We express
this condition with the natural inertial time associated with the characteristic force $F_0$: ($m$ is the mass of a disk of diameter $a$)
\be
T_0 = \sqrt{\frac{am}{F_0}},
\label{eq:defT0}
\ee
defining a dimensionless inertia parameter
\be
I_a = \dot\epsilon_{\text{max}}T_0.
\label{eq:defia}
\ee
$I_a$ is set to $0.05$ in our simulations. The main set of samples of types 1 and 2 (the latter coinciding with ``series A'' in paper I), to which some non-cohesive
ones are added for comparison, is listed in Table~\ref{tab:list},
in which the number of available configurations of different sizes is provided, along with
solid fraction under the lowest nonzero pressure. All configurations
are prepared both with ($\mu_R/a = 0.005$) and without ($\mu_R/a = 0$) RR, with the parameters of Table~\ref{tab:param}. The initial solid fraction is $\Phi_I=0.36$.
Type 2 systems are also available under $P^*=0$, right at the end of the aggregation stage~\cite{GRC07}, but we regard this intermediate stage as part of the
initial packing process and focus our study on higher pressures (as apparent in Table~\ref{tab:list}, the compression from zero pressure to $P^*=0.01$
involves a large density increase, and important changes of the microstructure are reported in paper I).
Distant interactions between grain pairs separated by a gap smaller than $D_0$ are scarce, and ``rattlers'', \emph{i.e.},
isolated, free  grains with no interactions, are absent in cohesive systems because of the initial aggregation process. Coordination numbers under $P^*=0.01$ are typically
$z\simeq 3.1$ without RR, and $z\simeq 3.0$ with RR, for both type 1 and type 2 cohesive samples. Additional details about those equilibrium configurations
under low pressure can be found in paper I.

The assembling stage of type 2 systems also depends on the initial velocities given to the
grains before they form aggregates (the ``granular temperature'' of the original ``granular gas''). The relevant dimensionless parameter is the ratio of the
initial mean quadratic velocity $V_0$ to the characteristic velocity $V^*$ defined in~\eqref{eq:vsep}. $V_0/V^*$ is set to $9.5$ for the main sample series of
Table~\ref{tab:list}. The value of $V_0/V^*$ was shown in paper I to have a strong influence on the initial coordination number $z$ at $P^*=0$ in samples
with RR: whereas $z$ is larger than 3 for $V_0/V^* = 100$, it approaches 2 for small $V_0$, of order $V^*/10$, in which case the loopless structures of geometric
ballistic aggregation models are retrieved. However, this effect is strongly reduced after the compression step to $P^*=0.01$.

In the following, unless otherwise specified, all results will pertain to the systems of Table~\ref{tab:list}, and measurements will be averaged over all available
samples, error bars on graphs extending to one sample to sample standard deviation on each side of the mean value.
\subsection{Simulation procedures \label{sec:proc}}
\subsubsection{Equilibrium conditions \label{sec:equil}}
One of the specificities of our simulations of cohesive packings under varying pressure is the approach, computing cost permitting, of the quasistatic
material response, in which all configurations remain close to mechanical equilibrium. Equilibrium conditions have to be stringent enough to enable an unambiguous
identification of the force-carrying contact network and a study of its elastic properties.
 Due to the frequent occurrence of small
contact force values, this requires forces to balance with
sufficient accuracy. We used similar criteria as in paper I, which, in agreement with other studies on cohesionless systems~\cite{RC02,iviso1},
were observed to provide adequately accurate force values. The tolerance levels on force and torque balance equations is expressed in terms of
a typical intergranular force value $F_1 = \text{max}(F_0,Pa)$. A configuration is deemed
\emph{equilibrated} when (1) the net force on each disk is lower than $10^{-5}F_1$; (2) the total moment on each disk is lower than $10^{-5}F_1a$;
(3) the difference between imposed and measured stresses is less than $10^{-5}F_1/a$; and (4) the kinetic energy per grain is less than
$5 \times 10^{-8}F_1a$. Those conditions being met, we could check that, in the absence of external perturbations (and of thermal motion), no remaining slow motion, creep
or aging phenomena were present in our systems: on waiting longer, only a very slow decrease of the remaining kinetic energy is observed. Furthermore, the computation of
the stiffness (or ``dynamical'') matrix, see Sec.~\ref{sec:compelas} provides an additional stability check.

\subsubsection{Compression\label{sec:comp}}
The sample series of Table~\ref{tab:list} are subjected to a stepwise compression cycle. In each compression step, external reduced pressure $P^*$ is multiplied
a constant factor $10^{1/8}\simeq 1.334$, and one waits until the new equilibrium configuration is reached, with the criteria stated in Sec.~\ref{sec:equil}.
A condition of maximum strain rate is enforced, in order to approach the quasistatic compression curve, as in the preparation process, on setting (see Eqs.~\ref{eq:defia} and
\ref{eq:defT0})
$I_a = 0.05$. Parameter $I_a$, on replacing, in its definition,
$F_0$ by the force scale $a P$ (in 2D) corresponding to the confining pressure is analogous to inertia parameter
$I$ used to assess dynamical effects in steady shear flow~\cite{RoRoWoNaCh06,Rognon08}, or
in the compression of non-cohesive granular packings~\cite{iviso1,iviso2}.
The compression program is pursued until $P^*$ reaches the maximum value $13.33$, above which negligible plastic collapse is observed. It should be noted that, thanks to the
high value of stiffness parameter $\kappa$ (see Sec.~\ref{sec:param}), the typical contact deflection $aP/K_N$ at this highest pressure level is still very small.
Then, the effect of decreasing $P^*$ back from its highest value to $0.01$ is also simulated. As no large structural changes occur on decompressing the system,
larger pressure jumps can be imposed on unloading.

The simulations are computationally costly, as in some pressure steps equilibration times of order $100\,T_0$ are required, while the time step for the integration of
the equations of motion is  a small fraction of ${\disty \sqrt{m/K_N} = T_0/\sqrt{\kappa}}$.
This limits the size and the number of samples, and the use of small strain rates. Some tests of
statistical significance and rate dependence of the results will be reported in Section~\ref{sec:macro}.
\subsubsection{Computation of elastic moduli\label{sec:compelas}}
We observe that once samples are equilibrated
according to the conditions of Section~\ref{sec:equil}, then the Coulomb
criterion $\abs{F_T} \le \mu F_N^e$, as well as the rolling friction
condition $\abs{\Gamma_R}\le \mu_R F_N^e$ are satisfied as \emph{strict}
inequalities in all contacts. No contact is ready to yield in
sliding, and with RR no contact is ready to yield in rolling
either.
This ensures that the response to small enough external load increments about a well-equilibrated state will be elastic and reversible.
Elastic moduli express elastic response, \emph{i.e.}, with no effect of tangential or
rotational sliding and no change in contact network topology and geometry.
To compute elastic moduli, we build the stiffness matrix $\ww{K}$ of the contact structure (also taking into account the distant interactions).
$\ww{K}$~\cite{GRC07} is a square matrix of order $3N+2$ (the number of degrees of freedom in the system), depending on
stiffness coefficients $K_N$ (replaced by $-F_0/D_0$ for the rare distant attractive bonds), $K_T$, $K_R$ (with RR), and on network geometry.
$\ww{K}$ is symmetric, positive definite (once the free translational motions of
the whole sample as one rigid body are eliminated) -- and thus the stability of equilibrium states is
checked. To compute elastic moduli, one solves a linear system of equations:
\be
\ww{K}\cdot {\bf U} = \fext
\label{eq:sysmod}
\ee
for the unknown displacement vector ${\bf U}$, containing all particle displacements and rotations, as well as strains
$\left(\epsilon _\alpha\right)_{\alpha=1,\, 2}$.
The right-hand-side of~\eqref{eq:sysmod} contains external forces and torques applied to the grains, which are set to zero, and stress increments
$\left(\Delta\sigma_\alpha\right)_{\alpha=1,\, 2}$ (the same procedure is followed in~\cite{SRSvHvS05} with 2D disk packings and in~\cite{iviso3} with 3D sphere packings).
On setting $\Delta\sigma_1 = 1$, $\Delta\sigma_2 = 0$, or vice-versa, one thus gets two separate measurements of the compliance matrix in our (statistically) isotropic systems,
from which moduli $C_{11}$ and $C_{12}$ are deduced, and hence the bulk modulus  $B = (C_{11}+C_{12})/2$ and the shear modulus $G=(C_{11}-C_{12})/2$.
\section{Material behavior under isotropic load\label{sec:macro}}

\subsection{Compression and pressure cycle with non-cohesive material\label{sec:NoCoh}}
Non-cohesive systems of Table~\ref{tab:list}, initially obtained by isotropic compression of a granular gas (like the
3D sphere packings of \emph{e.g.}, Refs.~\cite{iviso1} and~\cite{MGJS99}), are subjected to a compression cycle, in which reduced pressure
$P/K_N$ increases from its initial value $P_0/K_N=10^{-5}$, up to $P_1/K_N=1.33\times 10^{-3}$, and decreases back to $10^{-5}$.

\begin{figure}[htb]
\centering
\includegraphics[angle=270,width=8.5cm]{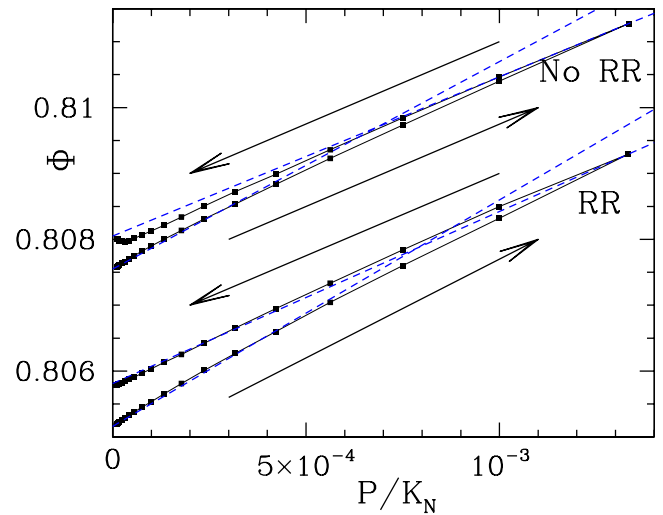}
\caption{(Color online) $\Phi$ versus $P/K_N$ in pressure cycle with 1400 disk samples with and without RR. Blue dashed lines
correspond to elastic response evaluated with the bulk modulus from initial and highest pressure states.
\label{fig:courbemodNC}}
\end{figure}

Typical results for the density of systems with and without RR are shown on Fig.~\ref{fig:courbemodNC}.
Changes of solid fraction
are very small (of order $10^{-3}$, i.e., of order $P/K_N$ for the largest pressure),
and nearly reversible (more than 90\% of the density increase is recovered on decompressing), as observed in Ref.~\cite{iviso2} with
3D sphere packings.
The slight increase of bulk modulus as a function of $\Phi$ is due to the larger density of contacts
under higher pressures.
One typical feature of frictional, cohesionless grain packs assembled by direct compression
is the existence of a non-negligible population of ``rattlers'', i.e., particles that transmit no force (as observed e.g. in Ref.~\cite{iviso1}
in 3D, or Ref.~\cite{SRSvHvS05} in 2D systems). The fraction of rattlers $x_0$
thus exceeds 20\% of the grains under $P_0$ in systems with RR in the
present case, and reaches 17\% without RR. $x_0$ is reduced to 14\% under $P/K_N=10^{-3}$.
The \emph{backbone} (force-carrying structure) is the set of non-rattler grains, characterized by coordination number $z^* = z/(1-x_0)$~\cite{iviso1}.
$z^*$ increases with $P$, as rattlers get captured by the backbone and gaps separating neighboring grains close in compression.

Changes of $x_0$ and $z^*$ are reversed on unloading (with some moderate hysteresis effect).
The increase of $z^*$ as a function of $P$, above a minimum value $z^*_0$, which would correspond to
$P=0$, is sometimes described by a power law~\cite{OSLN03}. With such a fit we can estimate  $z^*_0$, and we obtain
values close to 3 with RR
and about $3.12$ without RR. $z^*$ varies by about 10\% in the studied pressure interval.
As in other simulations~\cite{ZhMa05,iviso1,SvHESvS07},
the minimum coordination numbers stay above the ``critical'' value for rigidity, which is equal to 3 without RR and to 2 with RR~\cite{GRC07}.

Cohesionless systems under isotropic pressure cycles thus behave nearly elastically in an isotropic pressure cycle.
As the pressure increases by more than 2 orders of magnitude, while remaining in the rigid limit of $\kappa \gg 1$,
only small and nearly reversible changes in density and in other internal state variables are observed.
(see~\cite{iviso2} for a more detailed discussion).
A small level of RR has little effect on density and material properties.

\subsection{Compressing cohesive systems: general observations\label{sec:compgene}}
\begin{figure}[htb]
\centering
\includegraphics[width=8cm]{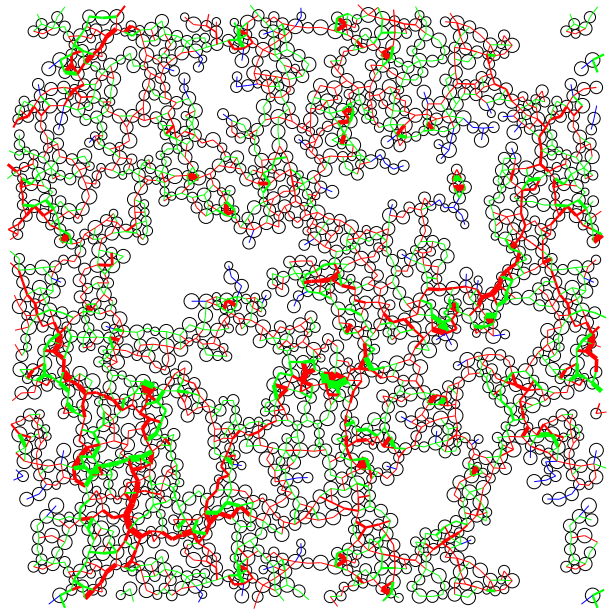}
\caption{\label{fig:config1} (Color online) Equilibrium configuration of a sample of 1400 disks with RR
in initial state, under $P^*=0.01$, for which $\Phi=0.5132$. Line thicknesses encode normal force intensities,
red strokes depict compressive forces while tensile ones are colored in green, and forces equal to zero in blue.}
\end{figure}

\begin{figure}[htb]
\centering
\includegraphics[width=8cm]{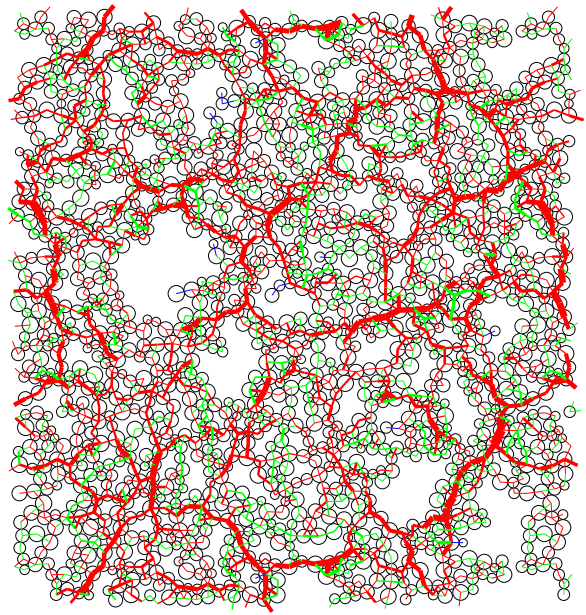}
\caption{\label{fig:config2} (Color online)
Sample of Fig.~\ref{fig:config1}, with $\Phi=0.6305$, equilibrated under $P^*=0.178$ (different length and force units).}
\end{figure}

\begin{figure}[htb]
\centering
\includegraphics[width=8cm]{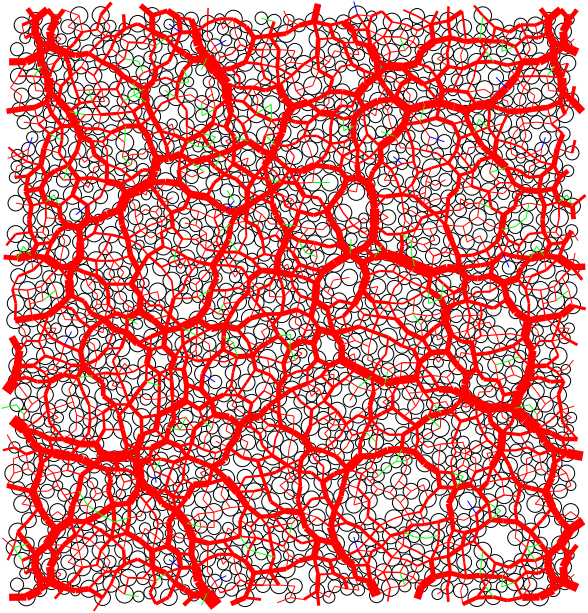}
\caption{\label{fig:config3}(Color online)
Same sample as on Figs.~\ref{fig:config1} and \ref{fig:config2}, under the maximum pressure $P^*=13.3$. Solid fraction is $\Phi=0.7778$.}
\end{figure}

Once subjected to a pressure cycle, as specified in Sec~\ref{sec:comp}, the material prepared in initially loose states (type 2 of Table~\ref{tab:list})
behaves as shown in Figs~\ref{fig:config1}, \ref{fig:config2} and~\ref{fig:config3}. As the pressure
increases, so does the density, and the large pores present under low $P^*$ gradually disappear.
The maximum packing fraction, $\phim = 0.774 \pm 0.001$ in that case, is quite
reproducible. $\phim$ is smaller than the solid fraction of cohesionless systems (for which $\Phi>0.805$,
see Fig.~\ref{fig:courbemodNC}).

From the shape of $\Phi(P^*)$ curves at growing $P^*$, three regimes can be distinguished.
At first, in a range of reduced pressure $P^*$ of the order of the first nonzero value ($10^{-2}$), thereafter called regime I,
$\Phi$ remain approximately constant: the contact network supports the growing pressure without
rearranging. Then, in a second pressure interval which we shall refer to as regime II, a fast
compression is observed. Density variations slow down in regime III, for $P^*$ of order unity, as a maximum solid fraction $\phim$ is approached.
On reducing the pressure, $\Phi$ then remains very close to $\phim$: the compaction is irreversible.

\begin{figure}[bht]
\centering
\includegraphics[width=8cm]{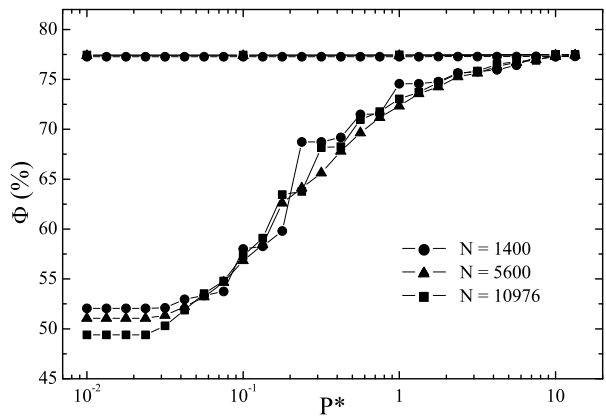}
\caption{Consolidation and decompression curves in 3 samples (with RR) with different numbers of grains, as indicated.
\label{fig:effN}}
\end{figure}

The consolidation curve is similar to the ones obtained by numerical simulations in Refs.~\cite{KBBW03,WoUnKaBr05},
on imposing uniaxial strains to loose packings prepared by an anisotropic ballistic aggregation process,
although our study differs from these works in several respects (see Section~\ref{sec:intro}). Refs.~\cite{KBBW03,WoUnKaBr05}
focus on regime III, and on dynamical compaction processes, with a shock wave propagating through the sample.
The variations of solid fraction $\Phi$ versus $P^*$ are shown
in Fig.~\ref{fig:effN}, for three samples of different sizes. Since all three curves are close to one another, we conclude that the macroscopic behavior is
correctly captured in our simulations.
Our results for $\Phi(P^*)$ also resemble experimental curves obtained on different materials, such as metallic powders~\cite{Poquillon02}, or
xerographic toner~\cite{MiguelAngel,C05}, at least in regimes I and II. Poquillon \emph{et al.}~\cite{Poquillon02},
in particular, in an experimental study of a metallic powder, explicitly distinguish three compaction regimes, with the material elastically
resisting compression in regime I, and then some plastic compaction, first attributed to particle rearrangement, as we observe, and later to
contact plasticity. This latter effect, which is not included in our model, is likely to explain the difference under high $P^*$ between
many experiments and our results: experimental curves do not appear to approach an asymptotic density, but witness ongoing compaction up to the
highest investigated pressure levels. In the case of metallic powders~\cite{Poquillon02}, quite high pressures are applied (hundreds of MPa), and,
as revealed by direct microscopic observations, particles fusing or sintering gradually form compact solids. For metal particles with d=10~$\mu$m
diameter, one can estimate the pressure $F_0/d^2$ corresponding to $P^*=1$ to be in the $0.1$~MPa range, so that the very large $P^*$ values
in the compaction experiment reveal a different physical origin of density increase.
The stiffness parameter, $\kappa$, is also significantly smaller in such experiments, with the consequence
that plastic phenomena cannot be ignored (for a definition and discussion of $\kappa$ in Hertzian sphere packings, see ~\cite{iviso2}).
Contact plasticity dominates in the numerical studies of
Martin~\emph{et al.}~\cite{MaBo03,Martin03,MBS03,Martin04}, which focus on
very high densities (beyond the random close packing value), when the material, due to sintering, turns into a porous compact. Hence only the early
stages of metal powder compaction, in which densities are quite low~\cite{Poquillon02} correspond to our simulations.
In the case of the xerographic toners studied in~\cite{MiguelAngel,CVQ05,C05}, $P^*=1$, as discussed in~\cite{GRC07},
rather correspond to $P\sim 10$~Pa. Nevertheless, the contact behavior,
as investigated by atomic force microscopy, is likely to involve plastic effects~\cite{Schaefer1994,Reitsma2000,QCV01,PhDFran}.

\subsection{Regime I: role of the initial assembling process\label{sec:regI}}
As shown in~\cite{GRC07} (paper I), and briefly recalled in Sec.~\ref{sec:init},
assembling conditions have a considerable influence on packing density and microstructure under low $P^*$. It should be assessed to what extent those important
differences in the initial configurations affect the plastic consolidation curve, and whether such a variability tends to disappear once the material undergoes
significant compaction. This issue is investigated in this section, in which the effects of various features of the preparation process are observed.
The role of some micromechanical parameters is also discussed.
\subsubsection{Compaction and aggregation in the assembling stage\label{sec:initM1M2}}
\begin{figure}[htb]
\centering
\includegraphics[width=8.5cm]{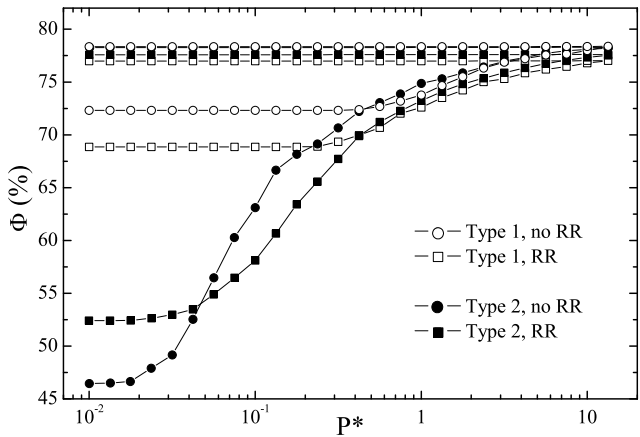}
\caption{Consolidation curve in type 1 and type 2 samples.
\label{fig:m1m2}}
\end{figure}

The most important feature of the assembling process is the competition between compression and aggregation, which leads to the difference between systems
of type 1 and 2, as defined in~\cite{GRC07} and recalled in Section~\ref{sec:init}. Type 1 samples reach a considerably higher densities
from the beginning, under low $P^*$. Fig.~\ref{fig:m1m2} compares the subsequent consolidation curves.
As type 1 systems are initially considerably denser, they are able to support larger pressures before rearranging, hence a wider regime I plateau.
However, the
pressure increase eventually reaches a high enough value to induce further compaction, and the consolidation curve is then very close to that
of type 2 systems (the difference is actually smaller than the sample to sample r.m.s. fluctuation). Within the accuracy and statistical uncertainty of our simulations, the
difference between initial states of types 1 and 2, although large, thus appears to disappear eventually upon plastically compacting the material.
\subsubsection{Effects of first compression step and strain rate~\label{sec:rate}}
\begin{figure}[htb]
\centering
\includegraphics[width=8.5cm]{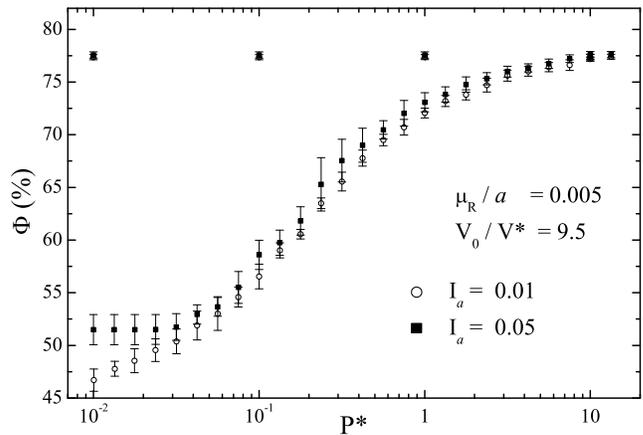}
\caption{Consolidation curve with two different values of $I_a$.
\label{fig:rate}}
\end{figure}

In paper I ~\cite{GRC07} important changes between $P^*=0$ and $P^*=0.01$ in type 2 configurations were reported, as solid fraction
$\Phi$ increases from $\Phi_I=0.36$ to about 0.5 (see Table~\ref{tab:list}).
One way to limit the effects of this first compression step
causes the most dramatic change is to reduce the strain rate, setting parameter $I_a$ to a lower value. As shown on Fig.~\ref{fig:rate}, displaying the consolidation
curve obtained in $N=1400$ systems with the usual value $I_a=0.05$ and with the smaller one $I_a=0.01$, lower inertial effects in the initial stage, while the
equilibrium configuration at $P^*=0.01$ is prepared, result in a lower density and tends to turn the initial plateau of the $\Phi(P^*)$ curve into a gentle ascending
slope. Later on, as consolidation proceeds, very similar curves are obtained with both values of maximum dimensionless strain rate $I_a$ (Fig.~\ref{fig:rate}), although
the smaller error bars (representing sample to sample r.m.s. fluctuations) witness smoother changes and better reproducibility for the slower compression. It may thus
be concluded that the quasistatic consolidation curve is quite reasonably approached with the standard compression procedure detailed in Section~\ref{sec:comp}, for which
$I_a=0.05$.
\subsubsection{Effect of initial agitation and influence of RR~\label{sec:agitRR}}
\begin{figure}[htb]
\centering
\includegraphics[width=8.5cm]{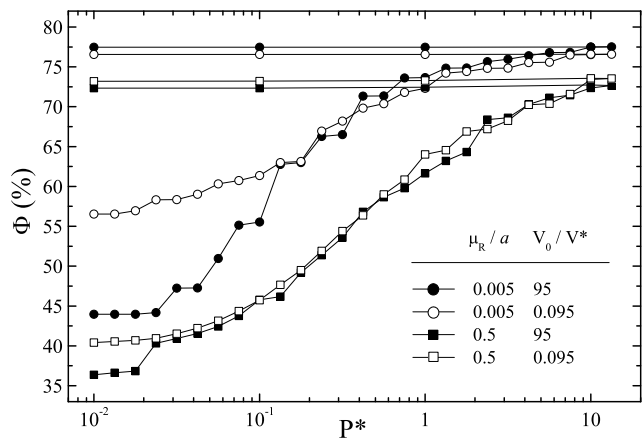}
\caption{Consolidation curve: effect of initial agitation level in aggregation stage, and influence of RR parameter.
\label{fig:V0RRphi}}
\end{figure}

\begin{figure}[htb]
\centering
\includegraphics[width=8.5cm]{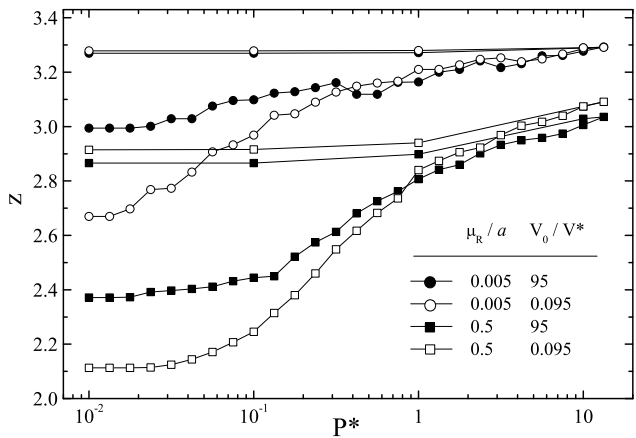}
\caption{Same as Fig.~\ref{fig:V0RRphi}, for coordination number $z$ as a function of $P^*$.
\label{fig:V0RRz}}
\end{figure}

The initial agitation velocity (or ``granular temperature''), as expressed by ratio $V_0/V^*$ in the aggregation stage
strongly influences the coordination number.
Figs.~\ref{fig:V0RRphi} and~\ref{fig:V0RRz} show how this initial influence affects the beginning of consolidation curves and, once again, fades out later on.
Consolidation curves are shown in Fig.~\ref{fig:V0RRphi} for two different values of $V_0/V^*$, one tenfold as large as the standard value $9.5$ used in
the sample series of Table~\ref{tab:list}, and the other one smaller by a factor of 100.
Fig.~\ref{fig:V0RRz} shows the effect of $V_0$ on coordination number.
An increase of rolling resistance (with $\mu_R = 0.5$ instead of $0.005$), similarly to a decrease of $V_0$, stabilizes looser systems under low $P^*$, with
smaller coordination numbers. However, such a change in material properties does not only affect the initial, regime I part of the consolidation curve; it
also alters the macroscopic mechanical behavior at larger densities: the slope of the consolidation curve is lower for larger RR.
\subsubsection{Conclusion on initial states and regime I}
Fragile tenuous structures due to aggregation are easily perturbed and sensitive to many factors in low consolidation states. In general, all perturbations
favor some kind of preconsolidation effect, inducing denser, better coordinated structures. These effects are reduced in each one of the following situations:
(1) if one waits until large aggregates form before applying a confining pressure; (2) if the initial agitation velocity
$V_0$ is decreased; (3) for slower compression processes, especially when the very first non-vanishing pressure value is imposed; (4) with larger RR levels.
As the material is further compressed in (nearly) quasistatic conditions,
the same macroscopic behavior is retrieved for given microscopic force laws [i.e., in cases (1) to (3)], irrespective of the initial perturbations affecting the beginning
of the consolidation process. Though we
did not vary the level of viscous dissipation in normal collisions,
lower values are expected to induce larger inertial effects, similarly to a faster compression. On the other hand, viscous forces slowing
down the motion of grains relatively to a surrounding fluid (often an important physical effect in fine powders) could reduce the effects of the initial agitation.

Regime I, with no plastic strain, is also observed in some experiments.
For example, the response in uniaxial compression (i.e., $\sigma_1>0$, $\sigma_2=\sigma_3=0$)
of loose aggregates of micrometer-sized silica beads assembled by ballistic deposition -- in that case, an anisotropic
process in which particles are thrown onto a substrate -- was studied by Blum and Schr\"apler~\cite{BlSch04}. The deposit, with volume fraction $\Phi\simeq 0.15$, resists
a stress of 500~Pa before plastic compaction is observed, which corresponds to a ``reduced stress'', defined, in analogy with $P^*$, as $\sigma_1^*\equiv \sigma_1 a^2 / F_0$
of order $10^{-2}$. In the simulations of Wolf~\emph{et al.}~\cite{WoUnKaBr05}
some finite initial pressure increment also has to be applied before plastic collapse is observed.
\subsection{Regimes II and III: \\
intrinsic consolidation behavior\label{sec:regII-III}}
Once the peculiarities of the sample preparation and first compression stage are erased, we refer to
the material evolution
as the \emph{intrinsic} consolidation behavior.
In order to compare the shape of the consolidation curve to other observations more directly and quantitatively (and also for
a more fundamental reason to be stated further) we subsequently describe it with $1/\Phi$, instead of $\Phi$, as a function of
$\log P^*$. This conforms to its traditional presentation in the literature~\cite{Wood90,BiHi93,Atkinson93,Poquillon02,C05}, which often uses the
\emph{void ratio}, $e=(1/\Phi)-1$.

Once the regime I ends, we obtain linear variations of $e$ or $1/\Phi$ with $\log P^*$:
\be
\frac{1}{\Phi} = \frac{1}{\Phi_0}-\lambda \ln \frac{P^*}{P^*_0}
\label{eq:consII}
\ee
%
where $P_0^*$ and the corresponding solid fraction $\Phi_0$ are the coordinates of the point where
the system behavior joins the intrinsic consolidation curve in the available samples.
Parameter $\lambda$, known as the plasticity index, is observed in our case to decrease as $\mu_R$ increases from zero (Fig.~\ref{fig:V0RRphi}).
We have also observed that the value of this index is not affected by the friction coefficient: in that sense, $\mu$ just displaces
the whole consolidation curve vertically~\cite{PhDFran}.

\begin{figure}[htb]
\centering
\includegraphics[width=8.5cm]{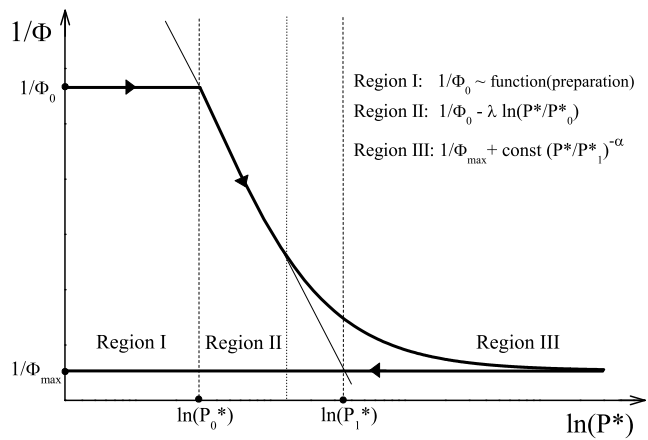}
\caption{Schematic view of intrinsic consolidation curve with regimes II and III, and role of parameters introduced in Eq.~\eqref{eq:cons}.
\label{fig:fit}}
\end{figure}

\begin{figure}[!htb]
\centering
\includegraphics[width=8.5cm]{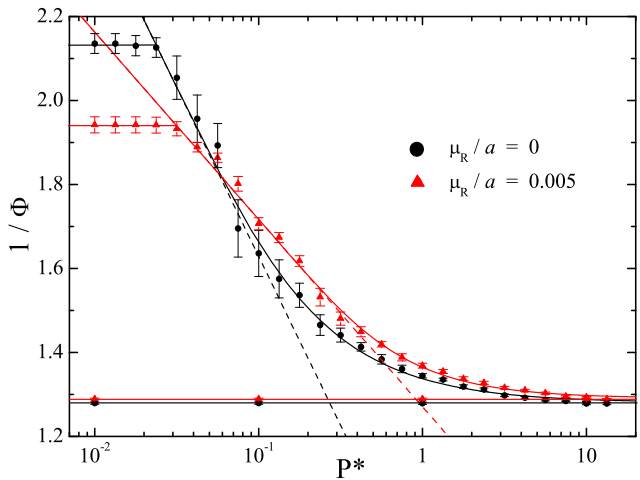}
\caption{(Color online) Consolidation data and fit to Eq.~\eqref{eq:cons}, for systems with and without (small) RR.
\label{fig:consfit}}
\end{figure}

As the maximum solid fraction $\phim$ is approached, Eq.~\eqref{eq:consII} is no longer valid, and the asymptotic regime is better described
with a power law, as in~\cite{WoUnKaBr05}:
\be
\frac{1}{\Phi} = \frac{1}{\phim} + \frac{A}{(P^*)^\alpha},
\label{eq:consIII}
\ee
with a constant $A$ and an exponent $\alpha$ (close to 1 in our results).
In order to describe the consolidation curve in regimes II and III with a unique functional form, we use the following relation:
 \be
\frac{1}{\Phi} = \frac{1}{\Phi_0}-\lambda
\ln \left\{ \frac{P^*}{P^*_0}\left[ 1-\exp\left( -\left[\frac{P_1^*}{P^*}\right]^\alpha\right) \right]^{1/\alpha} \right\},
\label{eq:cons}
\ee
which introduces additional parameters $P^*_1$ and $\alpha$, and crosses over from Eq.~\eqref{eq:consII}, for $P^*\ll P_1^*$, to
Eq.~\eqref{eq:consIII}, for $P^*\gg P_1^*$. Constant $A$ in~\eqref{eq:consIII} is set to $\lambda / (2\alpha)$ on using~\eqref{eq:cons} for large
$P^*$ values, and $P^*_1$ is directly related to $\phim$:
$$
\ln \frac{P^*_1}{P^*_0} = \frac{1}{\Phi_0}-\frac{1}{\phim}.
$$
Fig.~\ref{fig:fit} summarizes the definition and the role of all parameters of relation~\eqref{eq:cons}.
A fit of our data to relation~\eqref{eq:cons} is shown in Fig.~\ref{fig:consfit}.
It should be noted that even a small level of rolling resistance changes the plasticity index.
Values of parameters are listed in Table~\ref{tab:consfit}, where we also included the fit parameters for the
sample with $\mu_R/a=0.5$ corresponding to the data of Figs.~\ref{fig:V0RRphi} and \ref{fig:V0RRz}.
\begin{table}[htb]
\centering
\begin{tabular}{|c|ccccc|}
\hline
$\mu_R/a$ & $P_0^*$& $\Phi_0$& $\lambda$& $\phim$& $\alpha$\\
\hline
$0$& $0.0237$ & $0.469$& $0.349\pm 0.019$& $0.7808$& $0.91\pm 0.10 $\\
$0.005$& $0.0316$ & $0.515$& $0.194\pm 0.004$& $0.7745$& $1.08\pm 0.16 $\\
$0.5$& $0.0178$ &$0.382 $ &$0.25\pm 0.01 $&$0.724$&$0.86\pm 0.24$\\
\hline
\end{tabular}
\normalsize
\caption{Values of parameters $\lambda$, $\phim$ and $\alpha$ used to fit the consolidation curve
in systems of Table~\ref{tab:list}, and in a sample with larger RR, with Eq.~\eqref{eq:cons}.
Correspondingly, $P_1^*$ values are $0.271\pm 0.033$ without RR, $0.900\pm 0.064$ for $\mu_R/a=0.005$,
and $2.6 \pm 0.4$ for $\mu_R/a=0.5$.
\label{tab:consfit}}
\end{table}

As the consolidation curve in region II, defined by parameters $\lambda$ and $P_0^*$, is observed not to depend on initial conditions,  our simulations support
the following interpretation: sooner or later in
the process of quasistatic isotropic compression, the system joins, in the $P^*-\Phi$ plane, a certain locus, corresponding to compressive plastic yielding.
This locus, which acts as an attractor in isotropic compression, is a straight line on using coordinates $\ln P^*$ and $1/\Phi$.
The value of $P_0^*$  simply signals where, depending on the preparation process, the yield locus is reached.
Table~\ref{tab:param} gives the values of the parameters defining the intrinsic curve,
and of pressure $P_0^*$ where it is first reached in type 2 systems of Table~\ref{tab:list}.

Consequently, in a system prepared at a lower density, it should be possible to observe a wider interval of the intrinsic consolidation line.
We could explicitly check this property in the case of one sample with $N=5600$,
for which the first nonzero equilibrium confining pressure in the loading history
is equal to $2\times 10^{-3}$ instead of $10^{-2}$. This sample appears to have reached regime II sooner
(around $P_0^*=10^{-2}$, or possibly below). The corresponding data points lie on the
intrinsic consolidation curve (or, at least, within a distance smaller than error bars) identified on fitting the data
of the main sample series, which had a larger first compression step (to $P^*=10^{-2}$) and a larger value of $P_0^*$
(about $3\times 10^{-2}$). The yield locus can thus be \emph{extrapolated} to lower pressures and densities, with the same plasticity index $\lambda$.
\begin{figure}[!htb]
\centering
\includegraphics[width=8.5cm]{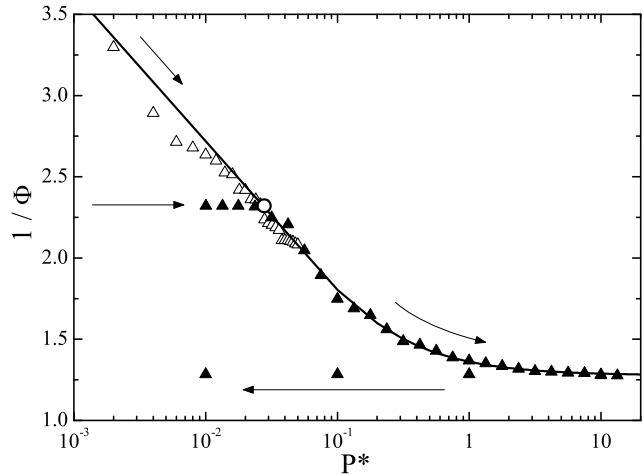}
\caption{Comparison of data obtained on the one low $P_0^*$ sample (open triangles), and Eq.~\eqref{eq:consII} (continuous line)
with the parameters of Table~\ref{tab:consfit}, as deduced from a fit of the data (black triangles) from the more systematic simulation series
with larger $P_0^*$.
\label{fig:lowP}}
\end{figure}
On assembling cohesive aggregates with arbitrarily low densities, and on stabilizing them
under very low initial pressures, it is conceivable (although increasingly difficult in numerical simulation because
of the computational cost, as well as in experiments, because of the system sensitivity to perturbations) to create equilibrium structures with smaller and
smaller densities and to explore an increasingly larger interval of the intrinsic consolidation curve in the limit of $P_0^*\to 0$.
The corresponding solid fraction $\Phi_0$ would then also tend to zero. This limit is compatible with the functional form used in Eq.~\eqref{eq:consII},
while the use of the alternative form~\cite{CVQ05,C05},
$$
\Phi-\Phi_0 = \nu\ln \frac{P^*}{P^*_0},
$$
would lead to contradictions in the limit of $P^*_0\to 0$.
\subsection{Unloading behavior\label{sec:unloading}}
\begin{figure}[!htb]
\centering
\includegraphics[width=8.5cm]{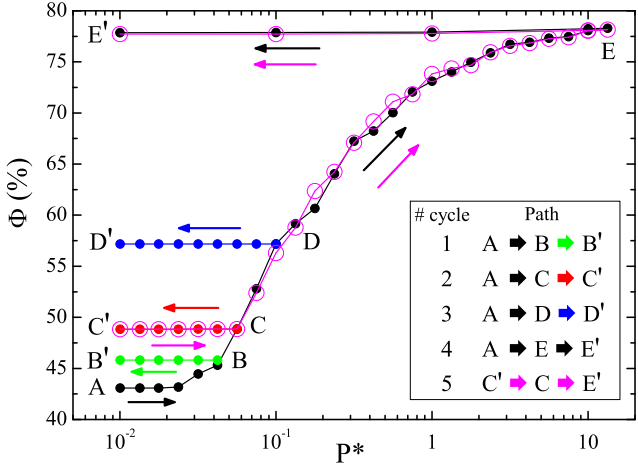}
\caption{(Color online)
Effect of different (isotropic) unloading/reloading histories on solid fraction. The direct consolidation curve with decompression from the highest pressure,
as shown in previous sections, is ABCDEE' (path 4). On unloading along lines BB', CC', DD', the system does not rearrange. Such paths are reversible and do not
alter the material state, since paths 4 (small black dots)  and 5 (large, open pink circles) superimpose in $P^*$, $\Phi$ plane.
\label{fig:reload}}
\end{figure}

\begin{figure}[!htb]
\centering
\includegraphics[angle=270,width=8.5cm]{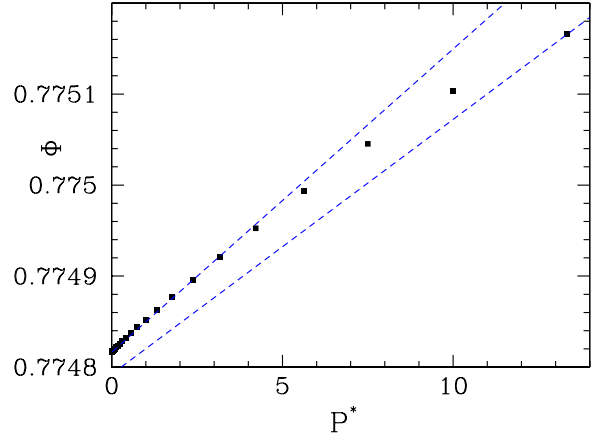}
\caption{(Color online)
Analog of Fig.~\ref{fig:courbemodNC}, for the unloading behavior of a sample with RR from $P^*=13.3$ to $P^*=0.01$. Dotted lines
correspond to the elastic response of the highest pressure state and the final state ($P^*=0.01$).
\label{fig:courbemodb04RR1}}
\end{figure}

On the $\Phi$ versus $P^*$ curves we have been showing so far, that the unloading branch, down to $P^*=0.01$,
shows very little density change.
This property is actually satisfied on decreasing the pressure from other configurations in
the compression process. Thus Fig.~\ref{fig:reload} shows that, if  $P^*$ is reduced to the initial level $0.01$ from different states on the consolidation curve,
density changes are hardly noticeable, and $\Phi$ stays very close to the maximum value reached at the largest imposed pressure
$P^*_c$ in the past. Furthermore,
it is checked (in the case of sequence 4, drawn with open circles in Fig.~\ref{fig:reload})
that the material might be reloaded, with no notable density change until pressure $P^*_c$ is reached.
$P^*_c$ is known in soil mechanics as the \emph{consolidation pressure}, and a material in a state such that $P^*<P^*_c$ is said to be
\emph{overconsolidated}.
Upon increasing the pressure beyond the consolidation value $P^*_c$,
the density irreversibly increases, and this compaction is described by the same curve as in
the absence of intermediate pressure cycle: the recompression curve from C' retraces back the same evolution from D to E. Thus the material behavior conforms to the
plasticity of clays in isotropic compression~\cite{Wood90}. All decompressing paths in the $P^*$, $\Phi$ plane, along which $P^*<P^*_c$, are reversible.
More precisely, they are similar to the pressure cycles applied to cohesionless systems (Fig.~\ref{fig:courbemodNC}), and they do not depart much from the
linear elastic response, as shown on Fig.~\ref{fig:courbemodb04RR1}.

For the largest $P^*$ values, adhesion forces are dominated by the confining stress
and are nearly negligible: on setting $F_0$ to zero in equilibrated systems under $P^*>10$, we could check that the granular assembly
finds a new equilibrium configuration with very small displacements and hardly any change in the contact network.

\section{Consolidation and density correlations\label{sec:geom}}
The gradual collapse of the initially open structure of loose systems, as visually apparent on Figs.~\ref{fig:config1}, \ref{fig:config2}, and \ref{fig:config3}
and witnessed by the consolidation curve studied in Section~\ref{sec:macro}, can be characterized by  the density correlation indicators introduced in paper I.

The initial aggregation process was shown in paper I to result in a fractal structure of the density field over intermediate scales, between the grain diameter and some
characteristic correlation length $\xi$. In the presence of rolling resistance, even with the small value $0.005a$ adopted for $\mu_R$, the observed fractal dimension
is compatible with the result of the ballistic aggregation model, $d_F\simeq 1.55$. The ballistic aggregation model is purely geometric, and corresponds to the
irreversible bonding of particles or aggregates in each collision, with contacts that are rigid in translation and rotation. This limit case, for which the
coordination number is equal to 2, is approached under
low pressure~\cite{GRC07} with large RR or small $V_0/V^*$. Better coordinated systems obtained with small RR and/or larger $V_0/V^*$ have the same fractal dimension.
Systems with no RR, on the other hand, are closer to dense objets with $d_F\simeq 1.9$~\cite{GRC07}.

The limitation of the fractal behavior by an upper length scale $\xi$ is a well-known
geometric necessity in a large system with finite particle packing fraction $\Phi$, because (in 2D) a fractal structure of dimension $d_F<2$
within a square cell of edge length $L$ exhibits an apparent density proportional to $L^{d_F-2}$. In physically relevant circumstances, systems with a finite packing
fraction $\Phi$ and a fractal structure over some distance range have a finite correlation length $\xi$ above which the average value of $\Phi$ is observed. One then has
$\Phi \propto \xi^{d_F-2}$ or
\be
\xi  \propto \Phi^{-1/(2-d_F)},
\label{eq:scalingphi}
\ee
the prefactor being specific to the particular system studied. Systems with size $L\gg\xi$
can then be regarded as homogeneous packings of fractal ``blobs'' of (linear) size $\xi$. Such ideas are quite generally used, and were applied to
semi-dilute polymer solutions~\cite{PdG79}, to silica~\cite{DACW86} or polymeric~\cite{COAD96} gels, in computer simulations of
aggregation models~\cite{MEA99}, and to various complex, supramolecular objects like
fat crystals~\cite{NaMa99} or asphaltene aggregates~\cite{RBD01}.

One may expect that the density increase caused by the collapse, under growing load, of the tenuous structures formed by cohesive packings corresponds to
a decrease in the fractal blob size $\xi$, while dimension $d_F$ still describes the scaling of density correlation at smaller scale. One should then observe the scaling
predicted in~\eqref{eq:scalingphi}. This implicitly assumes that the small scale structure of the packing is not affected by the compaction process, which essentially
breaks long, thin junctions and fills the largest pores. A clue in favor of such a scenario is provided by the results of Sec.~\ref{sec:regI}, which suggest
that the same structure is obtained if the material is directly prepared with some value of $\Phi$, or if it is assembled first in a looser state and then isotropically
compressed, up to solid fraction $\Phi$.

To compute
$d_F$ and $\xi$, we measure the ``scattering intensity'' $I(k)$, \emph{i.e.} the Fourier transform of the density autocorrelation function, as we briefly
recall now (see paper I for more details). Density field $\chi({\bf r})$, taking values 1 within particles and 0 outside, is first discretized on a regular mesh, then
Fourier transformed, thereby obtaining $\hat{\chi}(\kk)$. We then evaluate $I(\kk) = \abs{\hat{\chi}(\kk)}^2/A$, $A$ being the cell surface area. Invoking isotropy,
it is a function of $k=\norm{\kk}$ alone.
$I(k)$ should then vary proportionally to $k^{-d_F}$ for $a\ll 2\pi/k \ll \xi$, and reach some plateau for $k< 2\pi/\xi$.

This approach was used in paper I, and yielded the same fractal dimension, $d_F\simeq 1.52$ in systems with RR, under $P^*=0$ (solid fraction $\Phi_I = 0.36$) and
$\Phi=0.01$ (solid fraction $\Phi_0 =0.524\pm 0.008$), while $\xi$ decreased from
$\xi_I = 9.3\pm 0.4$ to $\xi_0 = 5.1\pm 0.2$. It should be noted that these values are roughly compatible with relation~\eqref{eq:scalingphi} (as
$(\xi_I/\xi_0)^{2-d_F} = 1.4\pm 0.1$ is close to $\Phi_0/\Phi_I= 1.46\pm 0.02 $).

\begin{figure}[htb]
\centering
\includegraphics[angle=0,width=8.2cm]{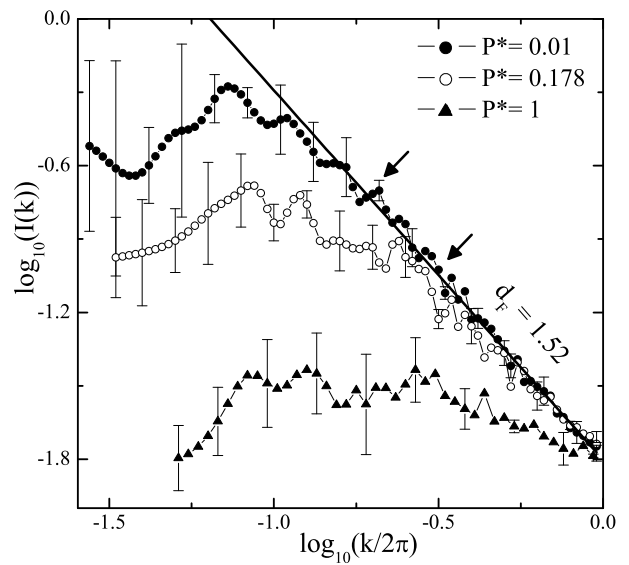}
\caption{Scattering intensity per unit area versus wave vector $k$.
Results are averaged over the four largest samples (with RR) of Table~\ref{tab:list}.}
\label{fig:Ik}
\end{figure}

Fig.~\ref{fig:Ik} shows the scattering function for similar consolidation states shown in Fig~\ref{fig:config1} ($P^*=0.01$), in Fig.~\ref{fig:config2} ($P^*=0.178$), and for $P^*=1$. These results are averaged over the four largest samples (with RR) of Table~\ref{tab:list}.
In spite of the error bars, $I(k)$ exhibits the expected form, it is approximately constant below some crossover wavevector $2\pi /\xi$ which increases with $\Phi$,
and then decreases, with slope  $-d_F$ on a logarithmic plot. Pressure $P^*=0.178$ is the largest one for which this latter feature is clearly observed, and
$I(k)$ data corresponding to smaller pressures are intermediate between $P^*=0.01$ and $P^*=0.178$ curves.
The arrows on the plot signal the identified values of wavevector $2\pi/\xi$, which values have been estimated by means of the fit function for $I(k)$
presented in paper I.
The curve corresponding to $P^*=1$ -- a flat, low
scattering signal -- is typical of dense, homogeneous media with no fractal range for density correlations.

In view of the small value of $\xi$ reached in the loosest configurations (those with $P^*=0$ studied in paper I),
relation~\eqref{eq:scalingphi} is difficult to test from density correlation data. Another characteristic length scale
for density inhomogeneities, used in paper I, is the (mass) averaged radius of gyration of pores.
It may provide an alternative definition of a blob size $\xi'$, proportional to $\xi$. We observed $\xi'\simeq \xi$ at $P^*=0.01$,
In fact, this equality works well under very low consolidations. However, under higher confining pressures we have observed that
the definition of $\xi'$ gives lower values than $\xi$.
\begin{figure}[!htb]
\centering
\includegraphics[angle=0,width=8.2cm]{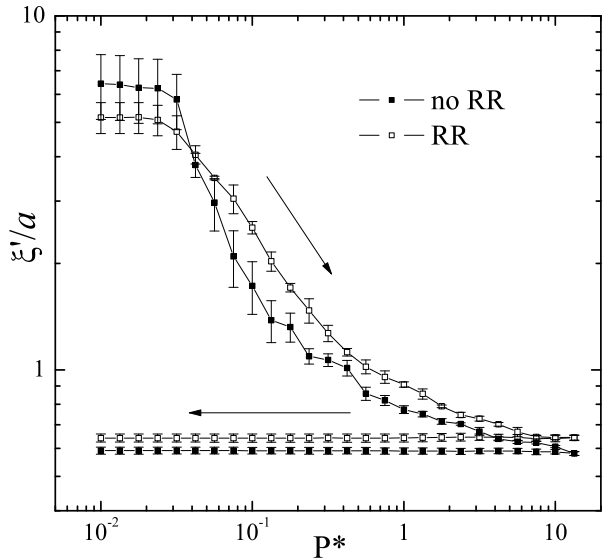}
\caption{Average radius of gyration of pores, $\xi'$, versus $P^*$.}
\label{fig:xi-P}
\end{figure}
Figure \ref{fig:xi-P} is a plot of $\xi'$ as a function of pressure.

Despite the restricted fractal range, our observations therefore confirm the validity of the ``fractal blob'' model, with a constant $d_F$ and a correlation length
$\xi$ decreasing as consolidation proceeds, until a final, homogeneous structure similar to that of cohesionless packings (albeit somewhat looser) is obtained. Other values of $d_F$ are likely to be observed
with other assembling processes (such as e.g., diffusion-limited cluster aggregation).

Values of $\xi$ and $d_F$ do not, however, entirely determine the mechanical properties of the system. The response of an aggregate to some mechanical perturbation should
depend on its connectivity which, as explicitly shown in paper I, is independent of its fractal dimension (systems with different $\mu_R$ and/or prepared with different
values of $V_0/V^*$ have the same $d_F$, but very different coordination numbers -- see also Section~\ref{sec:agitRR}).

Results concerning blob sizes in systems without RR, for which $d_F\simeq 1.9$~\cite{GRC07}, are similar.
Different stress states and
mechanical conditions might also produce other types of loose structures. As an example, in the slow steady state shear flow
of a very similar material simulated in~\cite{Rognon08} under normal reduced pressure $P^*=1.25\times 10^{-2}$ and shear stress $\sigma_{12} \simeq 1.5 P$ anisotropic structures with $\Phi\simeq 0.6$ were observed.
\section{Properties of equilibrium force networks\label{sec:forces}}
\subsection{Average normal force}
Formula~\eqref{eq:stress}, as explained in Ref.~\cite{RiEYRa06} and in paper I, leads to a simple relation between the average
normal force $\ave{F_N}$ in equilibrium, pressure $P$, solid fraction $\Phi$ and coordination number $z$:
\be
\ave{F_N}=\frac{\pi\ave{d^2}P}{z\Phi\ave{d}}=\frac{7\pi aP}{9z\Phi}.
\label{eq:rumpf0}
\ee
We observed formula~\eqref{eq:rumpf0} (involving the first and second moments of the diameter distribution)
to be accurate in all simulated states despite some approximations involved~\cite{GRC07}. However, as stressed in paper I, relation~\eqref{eq:rumpf0} fails to estimate the typical contact forces in the network under
low $P^*$. Those reach values of order $F_0$~\cite{RiRaEY06}.
Normal forces of both signs (as visible on Fig~\ref{fig:config1} and Fig~\ref{fig:config2}) coexist and, to a large extent, compensate under low $P^*$.
\subsection{Coordination numbers}
\begin{figure}[htb]
\centering
\includegraphics[angle=270,width=8.5cm]{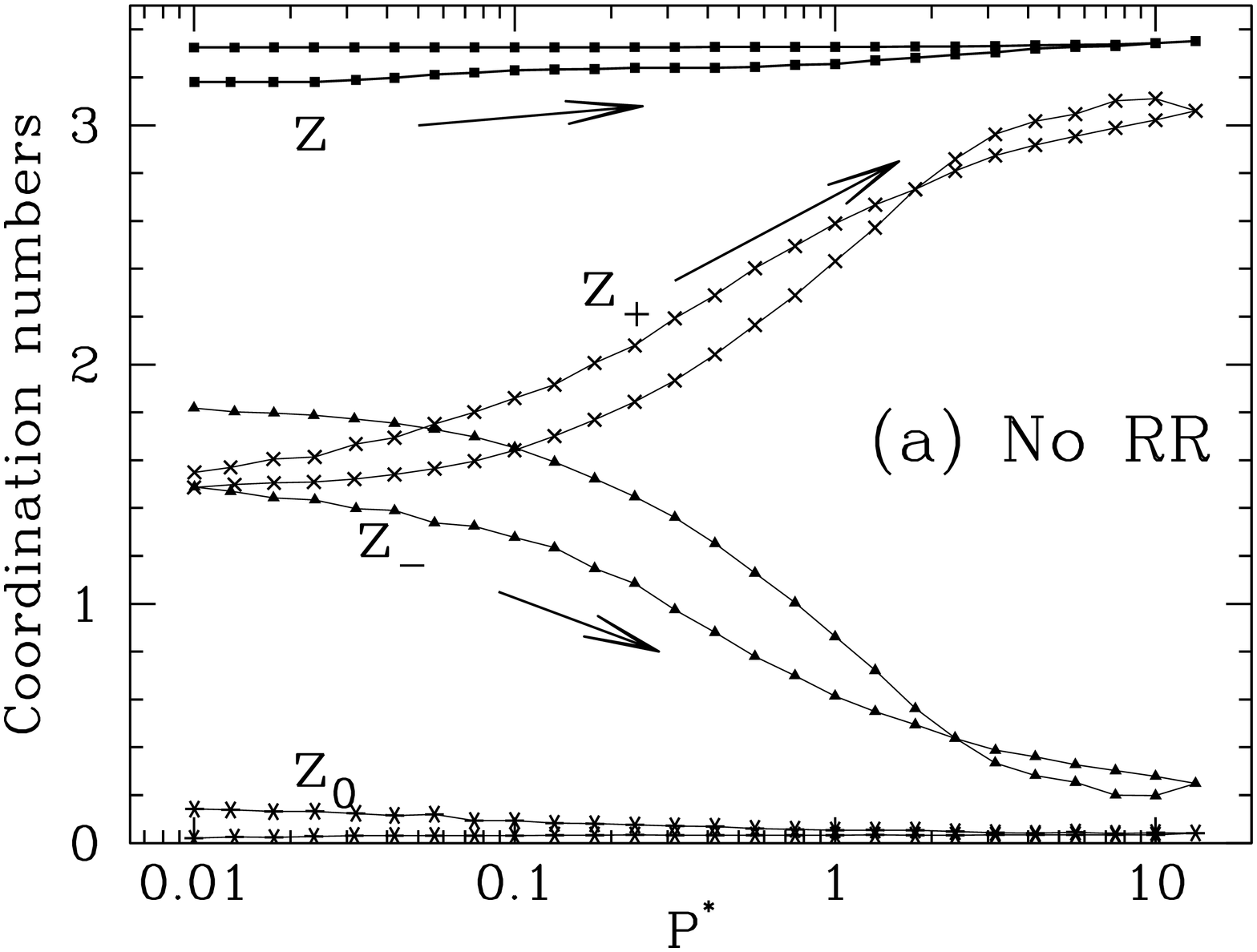}
\includegraphics[angle=270,width=8.5cm]{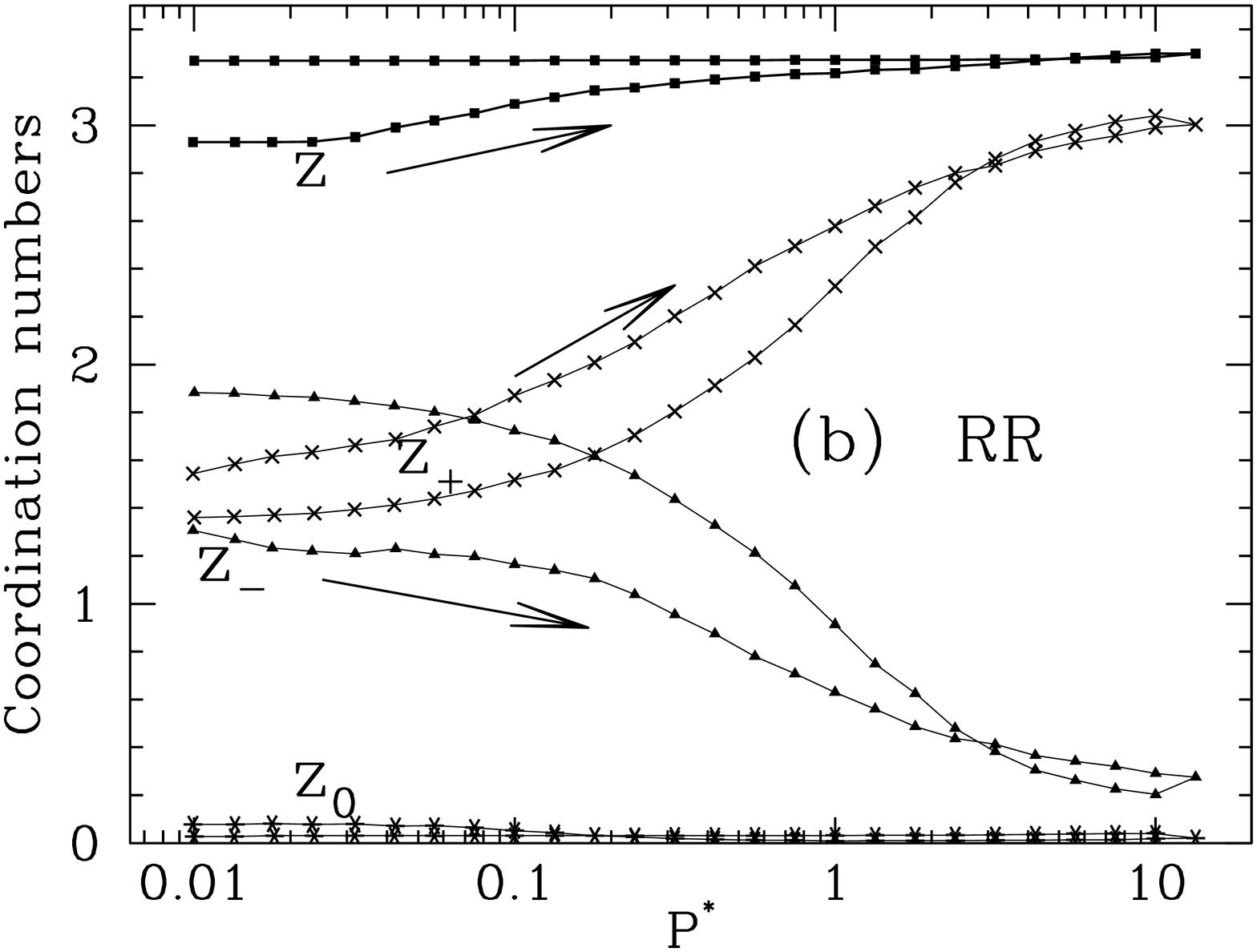}
\caption{\label{fig:coordh}
Coordination numbers versus $P^*$ in compression cycle (a) without and (b) with RR. Both plots display, from top to bottom,
$z$, $z_+$, $z_-$, and $z_0$. The error bars (not shown) are about the size of symbols. Arrows close to the curves indicate the compression branch of the pressure cycle.}
\end{figure}

In initial low-pressure states, the coordination number, $z$, as shown on Fig.~\ref{fig:coordh}, is nearly equally shared between
the contribution $z_+$ of compressive bonds and $z_-$ of tensile bonds. A small population ($z_0$ per grain) of contacts carry forces equal to zero (within the numerical
tolerance for force equilibrium).
Those contacts, in which the normal deflection $h$ takes the equilibrium value $h_0$ for isolated pairs~\cite{GRC07},
tend to be more numerous in the absence of applied stress, if the aggregation process avoids the building of hyperstatic (overbraced) structures. Their number is quickly
reduced once aggregates made under $P=0$ are subjected to some external stress and start rearranging.

The population of contacts loaded in compression increases along the consolidation curve until it dominates at large $P^*$. Upon unloading,
the initial proportion of tensile forces is first retrieved, and $z_-$ is eventually, under low $P^*$, larger than $z_+$.

The \emph{total} coordination number increases very little in the pressure cycle. Our observations thus
contradict some statements in the literature~\cite{JKL92,YZY03a} relating $z$ to $\Phi$ (even in cohesionless systems, $\Phi$ and $z$ can vary
independently~\cite{iviso1}).
%
%
In the course of
plastic collapse of loose structures, as the solid fraction increases by more than 50\%, we observe the number of contacts
to increase by 5\% in systems without RR, and by 12\% with RR. Such a small variation of $z$ in plastic compression contrasts with the comparatively very fast change
of $z$ in the quasielastic compression of cohesionless packings, as observed in Section~\ref{sec:NoCoh}, in which $z$ increases by more than 10\% for minute density
increases.

As to the number of distant attractive interactions, i.e. pairs of neighboring grains separated by a gap smaller than $D_0$ (contributing to $z_-$),
it is initially very low (typically 10 in a sample of
5600 particles), and then increases with $P^*$ but remains below 2\% of the total number of interactions.
\subsection{Distribution of forces}
\begin{figure}[htb]
\centering
\includegraphics[angle=270,width=8cm]{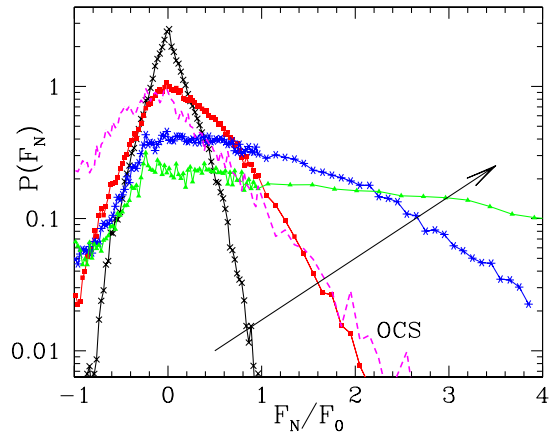}
\caption{\label{fig:histNR}
(Color online) Probability distribution function $P(F_N)$ of static normal force in contacts, versus $F_N/F_0$, in systems with no RR,
for $P^*=0.01$ (black),  $P^*=0.178$ (red), $P^*=1$ (blue), $P^*=2.37$ (green). Distribution widens as pressure increases as indicated by the arrow.
$P(F_N)$ is also shown for $P^*=0.01$ for the overconsolidated state (OCS) at the end of the pressure cycle (pink dashed line).}
\end{figure}

Normal force distributions are (roughly) symmetric about zero in initial states under low $P^*$~\cite{RiRaEY06}, as shown in Fig.~\ref{fig:histNR}.
Under low $P^*$, tangential forces of order $F_0$ are also frequently observed~\cite{GRC07}, and the angle between the total contact force
${\bf F}$ and the normal unit vector ${\bf n}$ is not constrained by the Coulomb condition, which
applies to  ${\bf F} + F_0 {\bf n}$ rather than to  ${\bf F}$. This explains the
typical patterns of self-balanced contact forces in small grain clusters, where compressive and tensile forces of order $F_0$ compensate locally,
as might be observed on Fig.~\ref{fig:config1}. The Coulomb
condition applying to ${\bf F}$, on the other hand, favors alignments and ``force chains''. Self-stressed small clusters form spontaneously when the disks aggregate,
except for large RR and/or small $V_0/V^*$~\cite{GRC07}.

\begin{figure}[htb]
\centering
\includegraphics[angle=270,width=8cm]{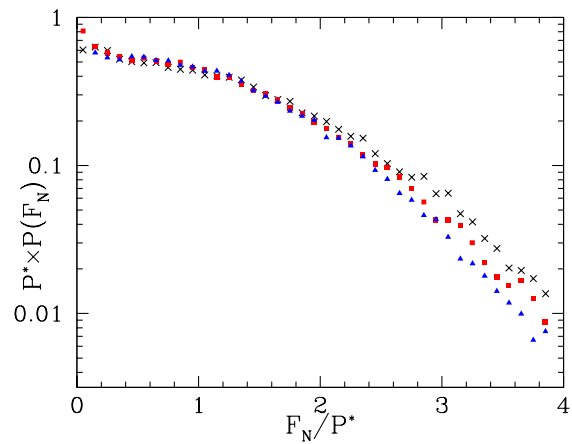}
\caption{\label{fig:histPNR}
(Color online) Positive wing of probability distribution function of rescaled normal
forces, $F_N/P^*$, in systems with no RR, under $P^*=2.37$ (black crosses),
 $P^*=5.62$ (red square dots), $P^*=13.3$ (blue triangles).}
\end{figure}

As consolidation proceeds, under growing $P^*$, normal force distributions develop a wider positive (compressive) side (Fig.~\ref{fig:histNR}),
while the finite value for $F_N=-F_0$ is characteristic of the failure of bonds in traction.
Forces eventually scale proportionally to $P^*$ at  large $P^*$, like in cohesionless systems~\cite{SRSvHvS05,iviso2}, as shown by Fig.~\ref{fig:histPNR}.
When $P^*$ reaches values of several unities, the force distribution is similar to that of cohesionless packings, with an additional dwindling population of
tensile contacts (Fig.~\ref{fig:coordh}). Force distributions in systems with small RR are quite similar to those shown in Figs.~\ref{fig:histNR} and \ref{fig:histPNR}.
\subsection{Forces in dense, overconsolidated states}
\begin{figure}
\centering
\includegraphics[width=8.5cm]{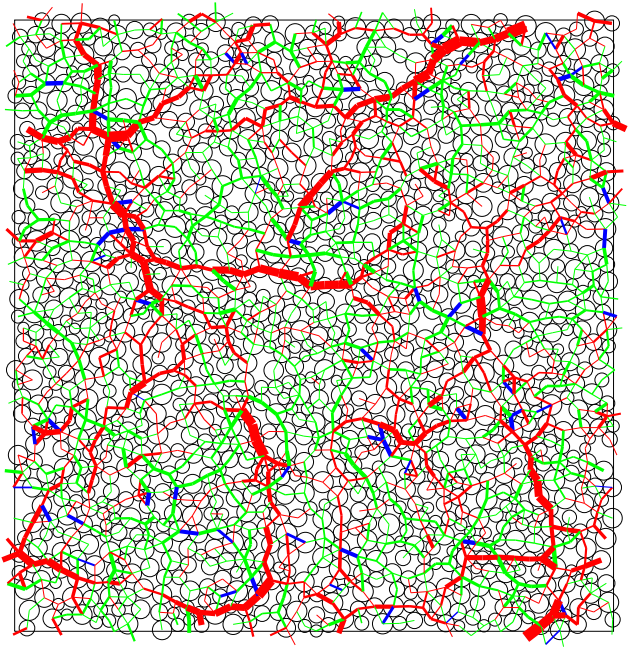}
\caption{\label{fig:overcon} (Color online)
Dense overconsolidated state of a sample under $P^*=0.01$ (with RR) at the end of the pressure cycle. Color code as on Fig.~\ref{fig:config1},
with distant attractive forces (for which $0<h<D_0$) in blue.
}
\end{figure}

Upon decompressing to low pressure levels, some larger compressive forces ($F_N/F_0$ reaching 2 or 3) survive and the distribution is not
symmetric (Fig.~\ref{fig:histNR}). Such effects of overconsolidation on contact forces
are considerably larger than in cohesionless granular materials~\cite{iviso2}. As in the case of cohesionless systems~\cite{iviso2}, we observed that
the decompression process tends to be affected by dynamical effects if it is too fast, and the overconsolidation effects on force distributions tend to be
erased if too many contacts open in transient stages. The results pertaining to overconsolidated states shown in Figs.~\ref{fig:histNR}, \ref{fig:coordh} and
\ref{fig:overcon} were obtained on simply reversing the stepwise compression program with the parameters indicated in Section~\ref{sec:comp} (i.e. with as many steps
in decompression as in compression).

This final force distribution is similar to the one reported  by Richefeu \emph{et al.}~\cite{RiRaEY06} in simulations of packings of wet
spherical beads, in which cohesion is due to capillary forces.
After assembling the packing under a finite pressure and then decompressing to $P=0$,
these authors observe that the particles tend to form small domains with only compressive or
only tensile forces. Fig.~\ref{fig:overcon} reveals quite similar patterns in overconsolidated states under $P^*=0.01$, with some
predominance of the regions under tension, while compressive forces tend to organize more often in strong force chains. Tensile contacts are more numerous than
compressive ones after the pressure cycle (Fig.~\ref{fig:coordh}).

To what extent overconsolidation effects on inner states influence the mechanical properties of cohesive granular materials (e.g., their response to shear stress)
would deserve to be investigated.

\subsection{Effect of a large rolling resistance\label{sec:RR}}
As reported in the previous paragraphs, the small level of rolling resistance used in most simulations reported here ($\mu_R/a= 0.005$)
has no large effect on force distributions or force patterns.
Yet, such a small RR significantly affects plasticity index $\lambda$ (see Table~\ref{tab:consfit})
and changes fractal dimension $d_F$ (Section~\ref{sec:geom}).

In order to understand the mechanisms by which RR affects macroscopic behavior and geometry, we investigated the effects of a large RR ($\mu_R/a=0.5$) in
a few 1400-disk configurations. Rolling resistance favors force transmission along thin strands of particles, each of them in contact with two neighbors (such structures are shown
in paper 1~\cite[Fig.~20]{GRC07}). Single particle chains are easily disrupted if $\mu_R/a$ is small, but are quite frequent for such RR levels.
While single particle chains are easily disrupted if $\mu_R/a$ is small, they become much more frequent for large rolling resistance
Thus coordination numbers may approach 2 (see Fig.~\ref{fig:V0RRz}). The density and the length of such particle chains are also witnessed by the
proportion $x_{22}$ of the contacts that join 2-coordinated
disks. Such contacts are impossible in an equilibrium structure without RR. $x_{22}$ reaches 12\% in large RR systems under low pressure
(for $\Phi$ in the 0.4 to 0.5 range), down to 1-2\% in the main sample series of Table~\ref{tab:list} with small RR ($\mu_R/a= 0.005$). 
Thin, rigid strands of 2-coordinated
disks might, however, be decorated by a side arm acting as a dead end for force transmission, and their mechanical role
is thus only partially captured on simply recording fraction $x_{22}$.
In the limit of $z\to 2$, which is approached under low pressure for large RR and/or low
velocity $V_0$ in the assembling stage, the force network has a vanishing number of loops and approaches isostaticity, as discussed in paper I~\cite{GRC07}. Consequently,
as compared to the case of small or no RR, systems with large RR under $P^*\ll 1$ exhibit narrower force distributions.
For $P^*$ of order $10^{-2}$, normal forces above $F_0/5$ or below $-F_0/10$ are extremely scarce (with
probability distribution function $P(F_N)$ in the $10^{-3}$ range). Furthermore, with $P^*\sim 1$, while compressive normal forces of order $F_0$ are frequently observed,
$P(F_N)$ remains below $10^{-2}$ for $F_N\to -F_0$. This contrasts with the results shown on Fig.~\ref{fig:histNR}: the proportion of contacts
on the verge of tensile rupture is much smaller in systems with large rolling resistance.
\section{Elastic moduli\label{sec:elasmod}}
Elastic moduli are used in experiments~\cite{KJ02}
and computer simulations~\cite{Makse04,iviso3} to express the response of granular materials to small
load increments. Their measurement, or that of
wave velocities, is a non-destructive probe of the packing structure. Thus, in the case of cohesionless bead packings, the simulations
of \cite{iviso3} showed that the moduli are sensitive to coordination number, which can vary independently of the solid fraction,
and escapes direct observations~\cite{iviso1}.
In the present case of possibly loose and poorly connected cohesive systems, those moduli also
approximately describe the parts of the compression curves with no packing rearrangement (Fig.~\ref{fig:courbemodb04RR1}), like
in cohesionless systems (Fig.~\ref{fig:courbemodNC}).

\subsection{Elastic moduli of cohesionless packings\label{sec:modulnocoh}}
\begin{figure}[htb]
\centering
\includegraphics[angle=270,width=8cm]{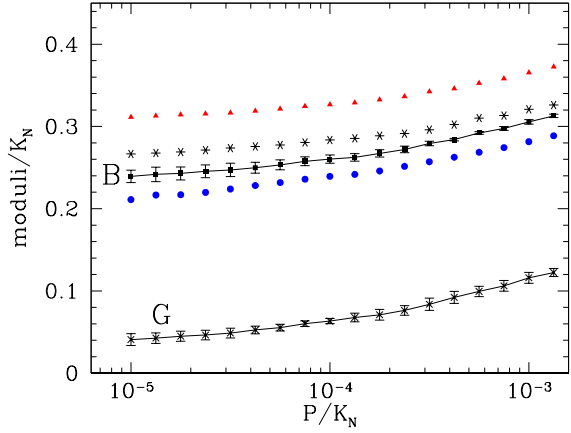}
\caption{(Color online) Bulk and shear moduli of cohesionless systems (with RR), versus pressure in compression cycle.
Voigt and Reuss bounds are shown as (red) triangles and (blue) round dots, respectively. Asterisks show values of $B$ obtained on taking a larger
rolling stiffness, $K_R = 10^{-2} K_Na^2$ instead of $K_R = 10^{-4} K_Na^2$, with the same contact network.
\label{fig:modNC}}
\end{figure}

We first quickly describe the variations of elastic moduli in the cohesionless systems of Table~\ref{tab:list}, and
their relations to microstructural or micromechanical parameters. Fig.~\ref{fig:modNC} is a plot of bulk and shear moduli
versus pressure. Values of moduli are very similar in systems without and with RR, and vary very slowly with $\mu_R$ in the latter case.
Unlike with Hertzian contacts, local stiffness constants $K_N$, $K_T$
do not depend on forces. Consequently, the increase of moduli with pressure is moderate.
The results of Fig.~\ref{fig:modNC} are typical of cohesionless granular systems
with small coordination number~\cite{SRSvHvS05,SvHESvS07,iviso3}. The
evolution of bulk modulus is correctly described by the simple estimation formulae
recalled below in Sec.~\ref{sec:modulestim},
and it is explained by the increase of coordination number. Shear modulus $G$, on
the other hand, is somewhat anomalously low,
witnessing the propensity of a rather poorly connected contact network ($z^*\simeq 3.1$ under
$P/K_N = 10^{-5}$, without RR) to rearrange under small stress increments, \emph{if those
are not proportional to the preexisting stresses}.

The evolution of elastic moduli in the unloading part of the pressure cycle (not shown on the figures, for clarity)
very nearly reverses the effect of the first compression.
\subsection{Simple estimation formulae\label{sec:modulestim}}
Bulk and shear moduli are traditionally estimated by the Voigt or mean field formula~\cite{WAK87,MGJS99},
which gives upper bounds~\cite{iviso3} $B^V$, $G^V$ in terms of contact stiffness constants and coordination number $z$, based
on the assumption that particle centers move like points of a homogeneously strained continuum. In the present case one has:
\be
\ba
B^V &=\frac{z\Phi \left[\ave{d^2}+\ave{d}^2\right]K_N}{4\pi\ave{d^2}} =\frac{55z\Phi K_N}{112\pi}\\
G^V &= \frac{K_N+K_T}{2K_N}B^V
\ea
\label{eq:bvgv}
\ee
On deriving~\eqref{eq:bvgv}, similar approximations are used as for~\eqref{eq:rumpf0}.
The formulae are identical for systems with or without RR, and since we chose $K_T=K_N$ one
has also $G^V  = B^V$.

For the bulk modulus, one may also write down a lower bound $B^R$, the Reuss estimate~\cite{iviso3}, based on the evaluation of the elastic
energy with trial forces in a load increment. The formula for $B^R$ involves moments of the contact force distribution, specifically the following ratio:
\be
\tilde Z_2 = \frac{\ave{F_N^2 + \frac{K_N}{K_T} F_T^2 + \frac{K_N}{K_R} \Gamma ^2}}{\ave{F_N}^2},
\label{eq:defZ2}
\ee
in which averages are taken over all contacts carrying static normal force $F_N$,
tangential force $F_T$ and rolling moment $\Gamma$ (to be set to zero in the absence of RR).
Using~\eqref{eq:rumpf0}, one has
\be
B^R = \frac{z\Phi \ave{d}^2 K_N}{2\pi\ave{d^2}\tilde Z_2}=\frac{27 z\Phi K_N}{56 \pi\tilde Z_2}.
\label{eq:br}
\ee
This approximation of the bulk modulus
becomes exact when the force increments caused by an isotropic pressure increase are proportional to the preexisting forces~\cite{iviso3}, and
hence it tends to be accurate in systems with small degrees of force indeterminacy.
The ratio of upper to lower bounds for $B$
given by Eqs.~\eqref{eq:bvgv} and~\eqref{eq:br} is $55\tilde Z_2/54$, and the bulk modulus
is therefore especially well predicted when the force distribution is not too wide~\cite{iviso3}, and ratio $\tilde Z_2$ stays close to 1.
Thus bulk moduli are rather successfully estimated (see Fig.~\ref{fig:modNC}) by $B^R$ or $B^V$ in the cohesionless case of
Sections~\ref{sec:NoCoh} and \ref{sec:modulnocoh}. Force distributions have often been studied in cohesionless systems, in which they are strongly constrained by
the no-tension condition, and $\tilde Z_2$ cannot reach large values ($\tilde Z_2 \le 1.5$ in the present case).
\subsection{Elastic moduli in cohesive packings\label{sec:modulcoh}}
\begin{figure}[tbh]
\centering
\includegraphics[angle=270,width=8cm]{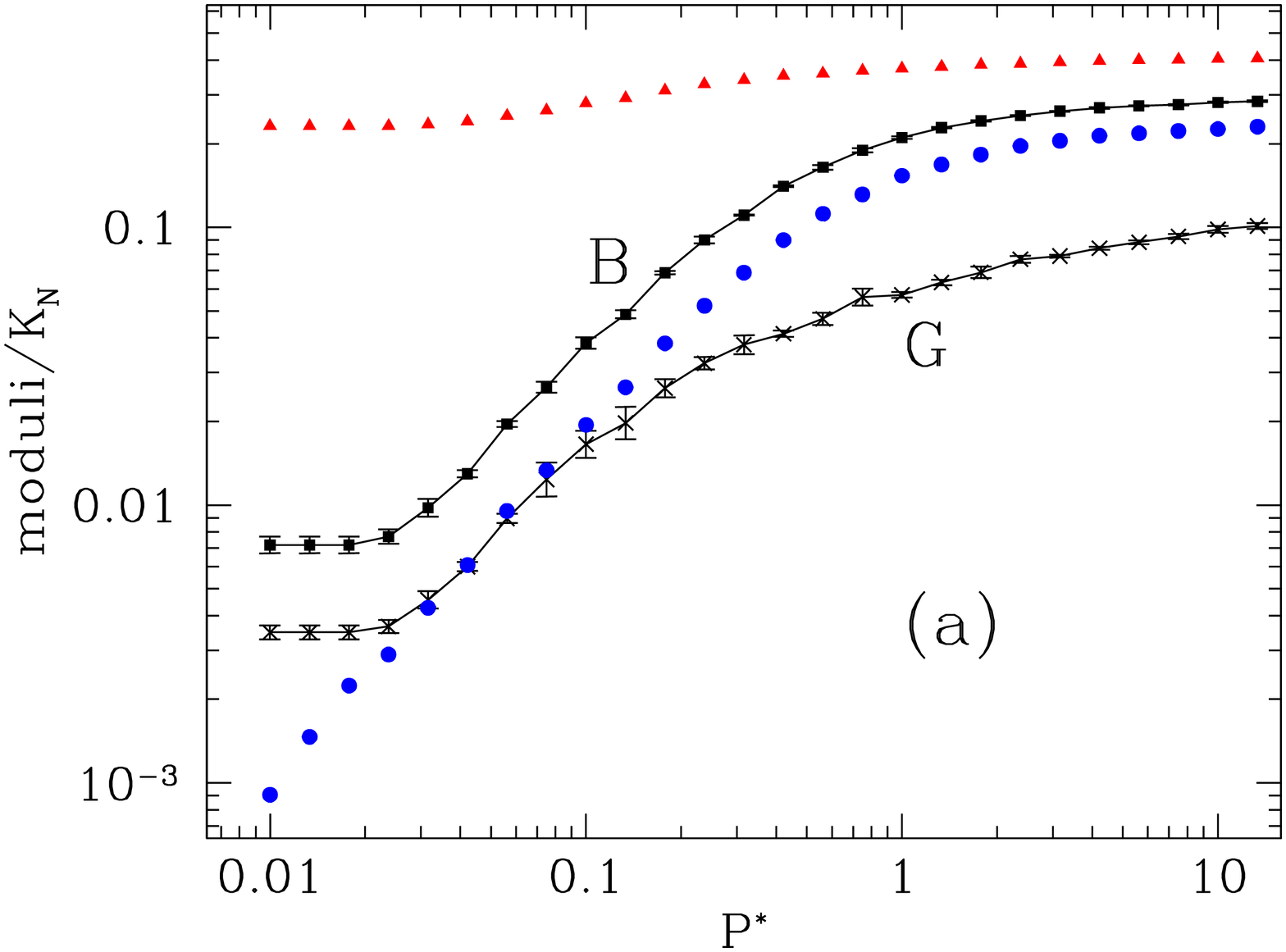}\\
\includegraphics[angle=270,width=8cm]{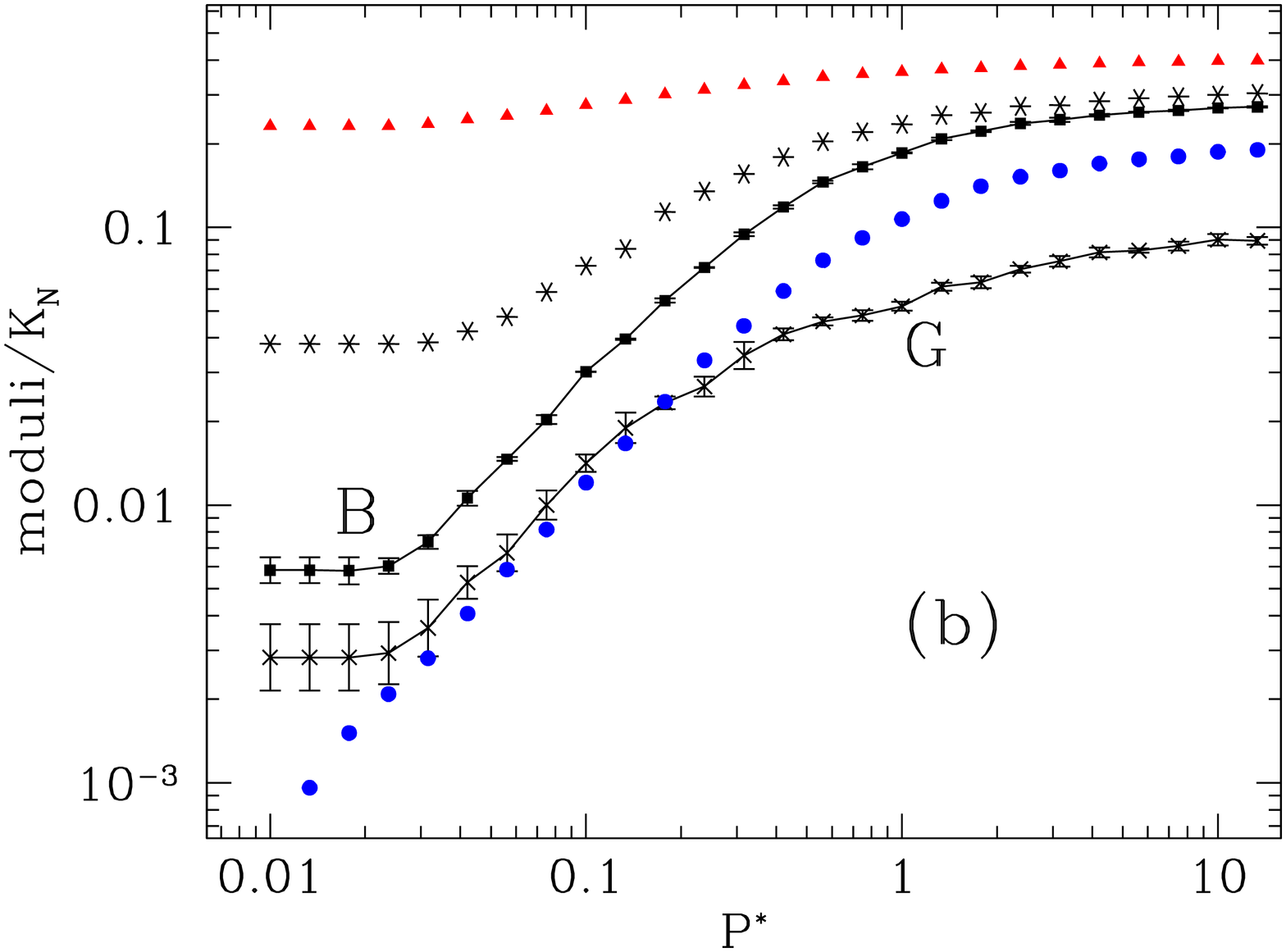}
\caption{(Color online) Bulk and shear moduli of cohesive systems (a) without and (b) with RR, versus (growing) pressure.
Same symbols and colors as in Fig.~\ref{fig:modNC}.
\label{fig:modmh}}
\end{figure}

Elastic moduli as functions of $P^*$ during consolidation of cohesive systems are plotted in Fig.~\ref{fig:modmh}.
Note the logarithmic scale used for elastic moduli (unlike in Fig.~\ref{fig:modNC}).
Both bulk and shear moduli are very low at small $P^*$,
which cannot be simply explained by the factor $z\Phi$ appearing in estimates~\eqref{eq:bvgv} and~\eqref{eq:br}
($z$ values, see Fig.~\ref{fig:coordh}, are similar to those of cohesionless systems
while $\Phi$ is twice as small at most). Those anomalously low moduli witness the propensity of the system to rearrange \emph{under
isotropic as well as under deviatoric stress increments}.
Moduli in samples with RR (Fig.~\ref{fig:modmh}b) have very similar values as in the absence of RR, although this may be partly coincidental, since they
are quite sensitive to the value of rolling stiffness $K_R$.

On decompressing, the moduli (not shown in Fig.~\ref{fig:modmh}) stay close to the value reached at the highest pressure.

Mean field estimates $B^V$ and $G^V$ are \emph{both} too large by factors of 30 to 50
in loose states. From~\eqref{eq:rumpf0} the average normal force $\ave{F_N}$ vanishes as $P^*$ tends to zero,
while the second moment is of order $F_0^2$. Moreover, as tangential forces are not limited by condition $\vert F_T\vert\le\mu F_N$,
but by $\vert F_T\vert\le \mu (F_N+F_0)$ instead, their contribution to the elastic energy is important (and so is
that of rolling moments in systems with RR).
Coefficients $\tilde Z_2$ thus reach values of order $10^2$ or $10^3$ under low pressure, whence $B^V/B^R\gg 1$
which is impossible in cohesionless systems.
The Reuss bound for $B$ is first (in regime I) too small by a large factor. Then, it seems to capture the evolution of the bulk
modulus in regimes II and III of the consolidation behavior. Ratio
$B/B^R$ is reduced to about 2 for $P^*$ of order $0.1$, and slightly decreases as compression proceeds.
It should be recalled, though, that the Reuss formula essentially relates the bulk modulus
to another unknown quantity, $\tilde Z_2$.
\subsection{Elastic moduli and force indeterminacy\label{sec:modhyper}}
\begin{figure}[htb]
\centering
\includegraphics[angle=270,width=7.5cm]{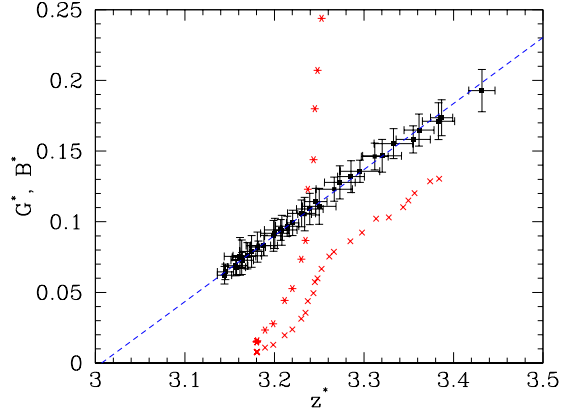}
\caption{\label{fig:modhyper} (Color online)
Elastic moduli (no RR) divided by $K_N\Phi(1-x_0)$, 
versus $z^*$, the coordination number without rattlers. Data with error bars, fitted with the dashed straight line,
correspond to $G$ in cohesionless systems. $G$ and $B$ in the cohesive material are respectively shown as (red) crosses and asterisks.
}
\end{figure}
The low value of the shear modulus in poorly coordinated cohesionless packings under isotropic stresses (see Fig.~\ref{fig:modNC}) has been observed~\cite{SvHESvS07,iviso3}
and argued~\cite{Wyart-th2} to stem from its tendency to vary proportionally to the degree of force indeterminacy per unit area (or volume in 3D) when it is small. As
the latter (without RR) is proportional to $(z^*-3)\Phi(1-x_0)$, one should have
\be
G^*\equiv \frac{G}{\Phi(1-x_0)}\propto z^*-3.
\label{eq:modhyper}
\ee
Fig.~\ref{fig:modhyper} shows our cohesionless packings to abide by this law, as the linear variation of $G^*$ with $z^*$ would predict, within uncertainties, its
vanishing for $z^*=3$.
However, it is also obvious from Fig.~\ref{fig:modhyper} that the anomalous
behavior of both moduli in loose, cohesive grain assemblies are not simply explained by their low coordination number, except perhaps for the
shear moduli of the densest configurations (rightmost data points), which, after sufficient plastic compaction, become similar to cohesionless packings.
A coordination number $z^*$ just above 3 (without RR) characterizes a ``barely rigid'' contact network, but such
a global, average quantity does not account for the specific heterogeneities of loose cohesive packings.
\subsection{Contact forces in a small pressure increment}
\begin{figure}[htb]
\centering
\includegraphics[width=8.5cm]{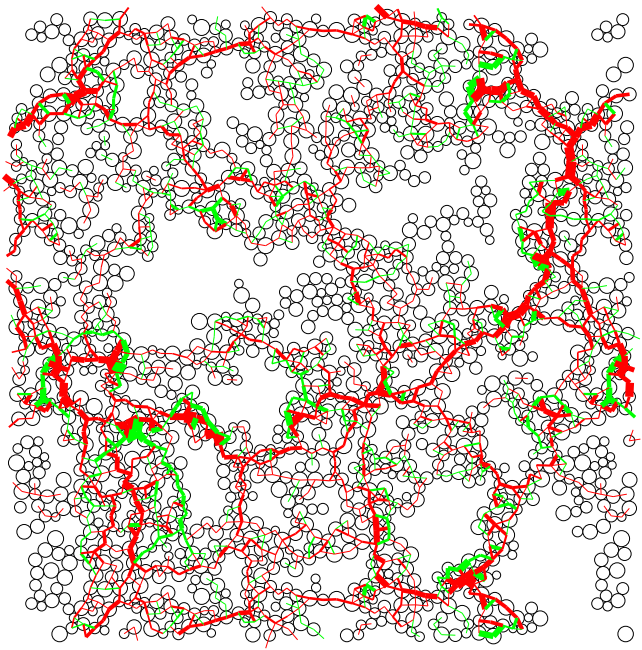}
\caption{\label{fig:AG3dP} (Color online) Force increments associated with elastic response in isotropic compression of system of Fig.~\ref{fig:config1}.
Contacts are ordered by decreasing contribution to elastic energy, and only the first 46\% contact forces corresponding to 95\% of the energy are drawn (colors as on
Figs.~\ref{fig:config1} to \ref{fig:config3}).
}
\end{figure}

At the microscopic level the elastic response to a small pressure increment $\Delta P$ determines contact
force increments as visualized in Fig.~\ref{fig:AG3dP}.
Very strong compressive force chains appear,
while large parts of the system carry very small
forces. On sorting the contacts by decreasing contribution to the elastic energy of the force increments balancing
$\Delta P$, less than half of them (46\%) contribute 95\% of the
energy. This proportion increases to about 65\% in the densest configurations, to be compared to 68--70\% in cohesionless systems. The configuration of
Fig.~\ref{fig:AG3dP}, in a system with RR, has quite a few dead ends, \emph{i.e.}, sets of grains that are connected to the rest of the structure but do not belong to any percolating
loop for force (or current) transport through the whole periodic cell. With RR, the force-carrying structure coincides with the backbone in the sense of ordinary (scalar) percolation
theory.
The force patterns of Fig.~\ref{fig:AG3dP} differ from those of Fig.~\ref{fig:config1}, in which the equilibrium forces, prior to the application of $\Delta P$ are shown: some
regions, especially the isolated, self-stressed clusters where compressions and tensions of order $F_0$ equilibrate, carry large forces but are bypassed in the
transmission of the pressure increment $\Delta P$. Dead ends contain ``islands'' of self-balanced forces resulting
from the aggregation process, as directly visible on Fig.~\ref{fig:config1}, but they do not participate in the transmission of stress increments and
they do not contribute to elastic moduli.

As consolidation proceeds, the repeated application of pressure increments clearly favors
force chains over localized self-stressed clusters, and the force pattern adapts to the external pressure.
Hence a closer similarity between the spatial distribution of
equilibrium forces under pressure $P$ and that of force increments caused by a small compression step $\Delta P$, and a better
performance of the Reuss estimate.
\subsection{Scaling with fractal blob size \label{sec:blobelastic}}
The inability of the approaches used in cohesionless systems to predict the elastic moduli of loose cohesive packings
can be attributed to their ignoring the peculiar network geometry,
which is the origin of the strong force concentration shown in Fig.~\ref{fig:AG3dP}.

In view of the results of Section~\ref{sec:geom}, it is tempting
to relate the elastic moduli to the variations of blob size $\xi$.
In scaling arguments about the density, the system can be regarded as a densely packed assembly
of somewhat fuzzy $\xi$-sized objects, the blobs.
To discuss elastic properties, the system is better represented as a network of ``superbonds'' of length $\xi$, or effective beams (with which the elongated structures
carrying stress in Fig.~\ref{fig:AG3dP} could be identified). In such a network, the dominant deformation mode is beam bending.
The transverse deflection $\delta$ in bending of a beam of length $\xi$, caused by a force $F$, is proportional to $\xi^3 F$.
Macroscopically, strains are of order $\epsilon  = \delta/\xi$,
while $F$ corresponds to stress $\sigma$ by $F\propto \sigma \xi$ in 2D. Consequently, the scaling of elastic moduli $\sigma/\epsilon$ with
length $\xi$ should involve a factor $\xi^{-3}$. (Some possible corrections to exponent 3 are possible, although the
appropriate value in, e.g., the case of percolation networks of beams is very close to 3~\cite{KaWe84,RouxS86}).
As $\xi$ varies by a factor of 3 or 4 within the scaling range (see Fig.~\ref{fig:xi-P}), relation $B\propto \xi^{-3}$
would predict an increase of moduli by a factor of a few tens.

Although this can be regarded as a fair estimate (see Fig.~\ref{fig:modmh}), it should be
admitted that the fractal range is very likely too restricted for such scaling laws to apply without important corrections.
With sufficiently large rolling resistance, the ``beams'' can be reduced to single particle chains, which, as we now show, enables simpler analyses
of their bending stiffness.
\subsection{The case of a large rolling resistance\label{sec:RRelas}}
With large RR the prevalence of particle strands as force-transmitting structures (Sec.~\ref{sec:RR}) influences elastic properties.
As noted above, linear structures tend to deform like bending beams,
with a compliance proportional to the third power of their length. In the case of single
linear strands, connections with contact properties are easily made more explicit. Consider, e.g.,
a straight, linear chain
of $n$ identical disks of radius $R$, with $n-1$ contacts characterized by stiffness constants $K_N$, $K_T$ and $K_R$. Then, in the elastic regime, all intermediate
disks can be suppressed and the interaction between the extreme ones, numbers 1 and $n$ along the chain, can be replaced by an effective one between two disks of radius
$(n-1)R$, and compliances $1/K_N^{(n)}$,  $1/K_T^{(n)}$, and $1/K_R^{(n)}$ for normal, tangential and rolling relative motion, with:
\be
 \left\{ \begin{array}{ll}
{\disty\frac{1}{K_N^{(n)}}}&= {\disty\frac{n-1}{K_N}}\\
{\huge \strut}{\disty\frac{1}{K_T^{(n)}}\strut}&={\disty \frac{n-1}{K_T}+\frac{(n-1)(4n^2-11n+6)R^2}{3K_R}}\\
{\huge \strut}{\disty\frac{1}{K_R^{(n)}}}&={\disty \frac{n-1}{K_R}}
        \end{array}
\right.
\label{eq:raidn}
\ee
For large $n$ the tangential compliance is much larger than the longitudinal and rolling ones, so that long chains behave as beams, which essentially deform
in bending. The local bending stiffness $EI$ of the beam (i.e. the product of the material Young modulus by the moment of inertia of the beam section) corresponding to
the chain of particles in the continuous limit is $EI=2RK_R$. (This coefficient expresses the proportionality of bending moment to rotation angle gradient).
For $n\gg 1$, the bending spring constant $3EI/l^3$ (expressing the transverse force to transverse deflection relationship)
is correctly identified from $K_T^{(n)}$ given in~\eqref{eq:raidn},
using the length $l=2(n-1) R$ of the straight $n$-particle strand.

Remarkably, the bending elasticity of small linear strands of micrometer-sized colloidal particles bound by adhesive forces has recently been measured by means of optical
tweezers~\cite{PaFu05}. Colloidal gels of polymer particles~\cite{PaFu04,FuPa07,PaFu08} should thus be modeled as cohesive particle assemblies with rather large RR level.

\begin{figure}[htb]
\centering
\includegraphics[angle=270,width=8cm]{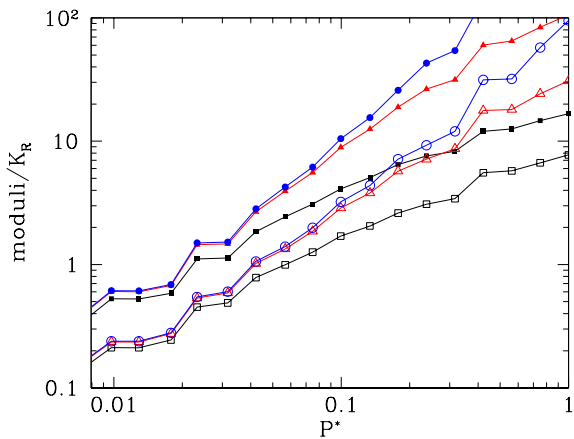}
\caption{\label{fig:modulW20}
(Color online) Bulk (filled symbols) and shear (open symbols) moduli, normalized by $K_R$, in low pressure states of a sample
for which $\mu_R /a = 0.05$ and $K_R = 10^{-2}K_Na^2$ (black
squares). Results obtained on evaluating moduli with $K_R = 10^{-3}K_Na^2$ and with $K_R = 10^{-4}K_Na^2$ are respectively shown as red triangles and blue circles.
}
\end{figure}

It is easy to check (consider \emph{e.g.,} two such chains joining at their ends at some
angle) that for all strand shapes other than straight lines, the extremities will be coupled by spring constants of order $K_T^{(n)}$ for both longitudinal (parallel
to end-to-end vector) and transverse relative displacements. Consequently, the macroscopic elastic moduli should be proportional to rolling stiffness constant $K_R$.
Fig.~\ref{fig:modulW20} shows that this proportionality is approximately satisfied in the loosest states of a system with rolling friction $\mu_R /a = 0.05$, in which
three different values of $K_R$ were used to evaluate the elastic response.
Elastic moduli of denser states, however, depart from this behavior. Therefore,
the scaling of elastic moduli with typical strand length (as suggested in Section~\ref{sec:blobelastic}) is
limited to low consolidation states.
With small or vanishing RR, single particle strands are replaced by thicker
junctions, which further restricts the consolidation pressure range for which elasticity is dominated by beam bending.

\section{Plastic consolidation mechanism: qualitative aspects\label{sec:compmicro}}
Cohesionless granular assemblies, if subjected to stress increments that are not proportional to initial stresses, essentially deform because the contact network
gets repeatedly broken and repaired~\cite{RC02,Staron02b}.
Macroscopic strains, once they exceed the very small scales associated with the response of given contact networks~\cite{RC02,iviso3}, thus result from a sequence of
rearrangement events or microscopic instabilities, during which the granular packing loses its coherence and gains some finite amount of kinetic energy, even for arbitrarily
slow applied stress changes. Collisions and appearance of new contacts stabilize the packing at the end of each microscopic rearranging event. This process gradually changes
the topology of the contact network, and produces specific evolution of its fabric (orientation anisotropy).

\begin{figure}[htb]
\centering
\includegraphics[angle=0,width=8.5cm]{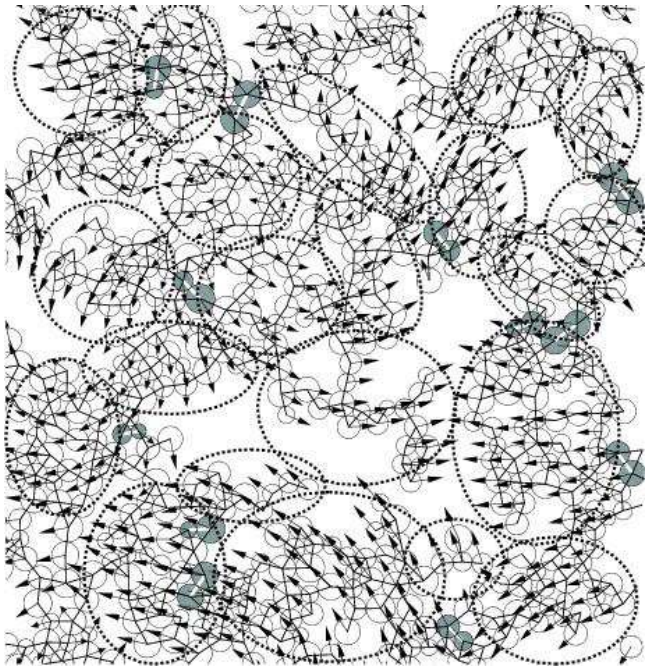}
\caption{Equilibrium particle positions in 1400 disk sample with small RR under $P^*=0.032$.
Particle displacements to new configuration equilibrated under $P^*=0.042$ are shown as arrows (global density change $\Delta\Phi = 0.05$).
Neighbor pairs for which contact opens are filled in grey. All other contacts (thin solid lines) are maintained.
Dense regions moving approximately like rigid solids are circled within dotted lines.
Most lost contacts are situated near the boundaries of such solid-like particle lumps.
}
\label{fig:mech-def-1}
\end{figure}

The mechanism of plastic collapse in isotropic compression of loose cohesive assemblies with small or vanishing RR in contacts, as observed in the present study, is similar.
Just like in cohesionless systems under shear~\cite{RC02}, we expect the frequency of occurrence of rearrangements, along the loading path, to increase, and the corresponding
strain jumps to decrease, as the size of samples grows, and thus the consolidation curve should be smooth in the thermodynamic limit. Due to the specific geometry of loose
systems, in which dense zones are weakly connected through thin arms, better connected, solid-like regions tend to move like rigid bodies,
while fragile junctions break and rearrange, so that initially large holes gradually fill up.
Fig.~\ref{fig:mech-def-1} illustrates this scenario. Displacements are depicted as arrows, pointing from the current positions to the ones reached in the next
equilibrium configuration in the stepwise compression sequence.
The more densely packed, nearly rigid regions (marked with dotted lines) are easily identified by direct visual inspection.
Fig.~\ref{fig:mech-def-1} also shows that the contact network undergoes relatively small topological changes, as more than
90\% of contacts are conserved. The rate of contact change, and the evolution of coordination number with strain are significantly smaller
than in cohesionless systems undergoing, e.g., shear deformation.
During the compaction of loose samples the dense regions collide and slide past one another, along thin sheared zones where most of the broken contacts are found.

In the case of large RR, the peculiar microstructure involving single particle chains might lead to a different deformation mechanism. Unlike multiply connected
junctions, simple strands can yield in bending without breaking: they fold at some contact, where the rolling friction threshold is reached, thereby releasing
bending elasticity. This mechanism is observed in experiments on single chains of colloidal particles~\cite{PaFu05,FuPa07}. One thus expects
fewer contact losses in plastic compression.

\begin{figure}
\centering
\vskip 7mm
\includegraphics[angle=270,width=8.5cm]{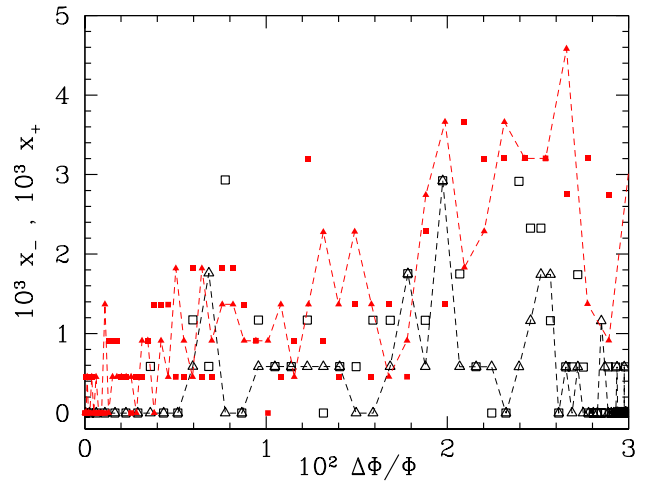}
\caption{(Color online) Evolution of the contact number as a function of relative density increase. In sample with $\mu_R/a = 0.005$ the proportions $x_+$ and $x_-$
of gained and of lost contacts with respect to the previous recorded list are respectively shown with red square dots and triangles -- the latter being connected with
a dashed line.   A similar code is used for  $x_+$ and $x_-$ values in a sample with large RR
($\mu_R/a = 0.5$), but with open dots, and in black.
}
\label{fig:mech-def-2}
\end{figure}

To follow more closely the rearrangement sequences in the course of compaction, it is appropriate to monitor changes in the list of contacts during
the motion between two equilibrium configurations.
As an example, let us consider the evolution between equilibrated states as $P^*$ increases from $0.177$ to $0.237$, and compare two samples, one with small
($\mu_R/a=0.005$) and the other with large ($\mu_R/a=0.5$) RR. Table~\ref{tab:complist} gives the changes in solid fraction and
coordination number, and numbers of maintained, destroyed and created contacts in this compression step.
Successive configurations separated by a fixed time interval $\Delta t=0.16 T_0$ are compared and
Fig.~\ref{fig:mech-def-2} plots the number of destroyed and created contacts as functions of time. For the same strain increment,
contact losses, as a function of global strain, are significantly less frequent in the sample with large RR.
This fact is reflected both in the data of Table~\ref{tab:complist}, where global changes are recorded, between the initial and final states,
and in those of Fig.~\ref{fig:mech-def-2}, where successive changes over time intervals $\Delta t$ are detailed.
\begin{table}[htb!]
\centering
\begin{tabular}{cccccc}
\hline
$\mu_R/a$ & $\Delta\Phi (\%)$ & $\Delta z (\%)$ & $N^{(=)}$        & $N^{(-)}$      & $N^{(+)}$      \\
\hline
0.005     & 3.2               & 0.14            & 2084 (94.9 $\%$) & 112 (5.1 $\%$) & 115 (5.2 $\%$) \\
0.5       & 3.1               & 1.2             & 1679 (98.5 $\%$) & 26 (1.5 $\%$)  & 46 (2.7 $\%$)  \\
\hline
\end{tabular}
\normalsize
\caption{Relative changes of solid fraction, $\Delta\Phi$, and of coordination number ($\Delta z$), and numbers of maintained ($N^{(=)}$),
destroyed ($N^{(-)}$) and created ($N^{(+)}$) contacts in a 1400 disks sample,
with small or large RR, in the compression step between $P^*=0.177$ and $P^*=0.237$.}
\label{tab:complist}
\end{table}
As a consequence, while the coordination number hardly changes during consolidation in systems with small or vanishing RR (see Fig.~\ref{fig:coordh}), it gradually increases
from an initial value close to 2 to nearly 3 in systems with large RR (Fig.~\ref{fig:V0RRz}). The lesser importance of tensile contact rupture in the plastic compression of
assemblies with large RR is also witnessed by the normal force distribution (Section~\ref{sec:RR}): forces approaching $-F_0$ are quite scarce, as opposed to the situation in
samples without RR (Fig.~\ref{fig:histNR}). With small RR, some single particle chains are also present, although shorter and less numerous. The  sensitivity of plasticity
index $\lambda$ to the rolling friction is likely to be
explained by different rupture mechanisms, the importance of folding rearrangements growing with the level of rolling resistance.
\section{Conclusion\label{sec:conc}}
To summarize, we have used  numerical simulations to observe and characterize, at the macroscopic and microstructural levels, the consolidation behavior,
in isotropic compression, of model cohesive powders. Macroscopic constitutives laws for quasistatic loading, unloading and elastic responses were shown to be reasonably well
approached. The material behavior was investigated for a range of densities that is wider than in most simulation studies of cohesive granular materials.
The consolidation process goes through three stages.
In a first regime, which is sensitive to the assembling procedure, no plastic collapse occurs, as the
agitation in the assembling process has stabilized a strong enough microstructure to withstand a finite pressure increase. The normal force distribution widens until a
significant fraction of contacts are on the verge of tensile rupture. The initial system geometry, which changes very little in regime I, is that of a dense assembly of
fractal blobs, with dimension $d_F$ taking the universal value associated with the aggregation process (here, ballistic) implemented in the sample preparation stage.
The blob size $\xi$ (at most, between 5 and 10 grain diameters in the present case) can be identified on studying density correlations.
The subsequent consolidation behavior is remarkably independent on initial conditions, which merely determine where the intrinsic consolidation curve in the $\Phi$-$P^*$ plane
is first met. The same curve is then followed whatever the initial conditions as the material is further compressed. This behavior corresponds, at the microscopic level, to
a gradual change of the blob size. The curve in regime II has the same shape as reported in the soil mechanics literature, and the consolidation pressure is a plastic
threshold below which the material response is approximately elastic (like the behavior of a cohesionless granular material under isotropic load). Elastic moduli
increase rapidly with consolidation pressure or density. The cases of small RR or without RR should be distinguished from the situation of strong rolling resistance, although,
in both cases, the microstructure of loose packings might be viewed as denser, better connected regions joined by thin arms.
In the first case, loose packings collapse when the tensile strength of contacts is overcome by the externally imposed forces, preferentially within the fragile
junctions between adjacent denser blobs. Systems with strong RR, on the other hand, contain single particle strands, which tend to fold without breaking in plastic compaction.
While small RR systems gain very few contacts in the consolidation process, the coordination number might increase from nearly 2 to 3 with large RR.
Eventually, the material approaches a limiting, maximum
density (regime III), as the packing structure resembles that of a cohesionless system, for $P^*\gg 1$ (albeit, typically, somewhat looser).
The absence of a similar upper limit of the density of
cohesive packings in experiments for large $P^*$ is due to plastic deformation of contacts.

The fractal blob size $\xi$, depending on solid fraction $\Phi$, is a central microstructural feature, based on which some scaling laws for elastic properties can be attempted.
It is also tempting, beyond the qualitative description of the microstructural changes associated with the consolidation process, to try to predict
the consolidation curve from such geometric data. Yet scaling laws only apply  to a restricted part of the consolidation pressure interval.

Our results, in many respects, emphasize important qualitative differences between cohesive and cohesionless granular assemblies. The existence of stable loose
structures and the consolidation phenomenon are the most important differences in macroscopic behavior brought about by cohesion. At the microstructural level,
unlike in cohesionless packings, the typical values of intergranular forces, or the force distribution,  are not as simply estimated in cohesive systems, in which
attractive and repulsive contact forces of the order of tensile strength $F_0$ tend to compensate under low pressure. In particular, compression cycles
stabilize self-balanced force networks with large compression forces.  Unlike in granular packings devoid of cohesion, the coordination number does not appear to be
a significant state variable in cohesive systems with low RR, as it hardly changes along the consolidation curve. With large rolling resistance, it witnesses, however,
the formation of loops under compression. While cohesionless assemblies with low coordination number usually contain many rattlers, all particles in cohesive packings are
connected to the same contact structure, which is rigid, but comprises lots of  ``dead ends'' or ``side arms'', which might bear self-balanced forces
but do not participate in the transmission of external stresses. Some of these new features can be summed up on remarking that loose powders are similar
to gels as much as to granular packings with no cohesion.

Our investigations should be pursued in several directions. On the theoretical side, the connections between macroscopic properties and microstructure could be studied
more quantitatively. The behavior of loose cohesive packings under general stress states should be investigated. Thus one may determine whether such constitutive laws as the
Cam-clay model~\cite{Wood90} apply to the simulated material.
And finally, more quantitative agreement with experiments and real materials
should be sought. In spite of some obvious steps (e.g., one should simulate 3D systems), this latter objective looks daunting.
One major difficulty is the importance of hydrodynamic effects at the assembling stage,
when the microstructure and the fractal dimension of aggregates are determined. While we have bypassed this problem on implementing ballistic aggregation, it is
necessary to investigate the behavior other possible kinds of aggregates, by dealing with some tractable model for hydrodynamic forces. It is hopefully
possible to introduce some mechanics and intergranular interactions within the models used with geometric aggregation rules (such as, e.g.,
diffusion-limited cluster-cluster aggregation).
Then, another  difficulty is that many parameters
associated with the  contact law  (such as friction coefficient, rolling friction, rolling stiffness constant) should be identified for a real material to be
investigated at the grain level. In this respect, the recent progress of experimental methods of microscopic investigation seems quite promising, as formerly
inaccessible parameters ruling interparticle contact mechanics are now beginning to be measured in model materials,
thanks to particle-scale observation and micromanipulation techniques~\cite{Schaefer1994,Reitsma2000,PaFu05,FuPa07,HeBuBlSch08}.

\begin{acknowledgments}

This work has been supported by the Ministerio de Educaci\'on y Ciencia
of the Spanish Government under project FIS2006-03645
and by the Junta de Andaluc\'ia under project FQM-421.

\end{acknowledgments}

\end{document}